\documentclass[11pt]{article}
\usepackage[margin=1in]{geometry}
\usepackage{amsfonts, amsmath, amssymb}
\usepackage[none]{hyphenat}
\usepackage{bm} 
\usepackage{siunitx} 
\usepackage[english]{babel}
\usepackage{graphicx}
\usepackage{url}
\usepackage{enumerate}
\usepackage{booktabs,multirow}
\usepackage{fixltx2e}
\usepackage[labelsep=period]{caption} 
\usepackage{hyperref}
\usepackage{subfig} 
\usepackage{sidecap} 
\usepackage{floatrow} 
\usepackage{float} 
\usepackage[affil-it]{authblk} 
\floatstyle{plaintop}
\restylefloat{table} 
\usepackage[titletoc]{appendix}
\usepackage{stmaryrd} 
\usepackage{placeins} 
\newcommand\keywords[1]%
  {\begin{flushleft}
   \let\and\\%
   \textbf{Keywords:}
   #1
   \end{flushleft}%
  }
  
\begin{document}
\sloppy 
\title{An immersed interface-lattice Boltzmann method for fluid-structure interaction}
\author[1,2,*]{Jianhua Qin}
\author[2]{Ebrahim M. Kolahdouz}
\author[3,**]{Boyce E. Griffith}

\renewcommand\Affilfont{\fontsize{9}{10.8}\itshape}
\affil[1]{Key Laboratory of Transit Physics, 
Nanjing University of Science  and Technology, Nanjing, China}
\affil[2]{Department of Mathematics, University of North Carolina, Chapel Hill, NC, USA}
\affil[3]{Departments of Mathematics, Applied Physical Sciences, and 
Biomedical Engineering, \protect\\University of North Carolina, Chapel Hill, NC, USA}
\affil[*]{Corresponding author at: Key Laboratory of Transit Physics, Nanjing University of Science and Technology, Nanjing, China.}
\affil[**]{Corresponding author at: Departments of Mathematics, Applied Physical Sciences, and Biomedical Engineering,
University of North Carolina, Chapel Hill, NC, USA.}
\affil[•]{Email: njustqin@gmail.com (J. Qin), boyceg@email.unc.edu (B. E. Griffith)}

\date{\vspace{-5ex}}
\maketitle

\begin{abstract}
An immersed interface-lattice Boltzmann method (II-LBM) is developed for modelling fluid-structure systems.~The key element of this approach is the determination of the jump conditions that are satisfied by the distribution functions within the framework of the lattice Boltzmann method when forces are imposed along a surface immersed in an incompressible fluid.  
In this initial II-LBM, the discontinuity related to the normal portion of the interfacial force is sharply resolved by imposing the relevant jump conditions using an approach that is 
analogous to imposing the corresponding pressure jump condition in the incompressible Navier-Stokes equations.
We show that the jump conditions for the distribution functions are the same in both single-relaxation-time and multi-relaxation-time LBM formulations. 
Tangential forces are treated using the immersed boundary-lattice Boltzmann method (IB-LBM). The performance of the II-LBM method is compared to both the direct forcing IB-LBM for rigid-body fluid-structure interaction, and the classical IB-LBM for elastic interfaces.
Higher order accuracy is observed with the II-LBM as compared to the IB-LBM for selected benchmark problems. Because the jump conditions of the distribution function also satisfy the continuity of the velocity field across the interface, the error in the velocity field is much smaller for the II-LBM than the IB-LBM.
The II-LBM is also demonstrated to provide superior volume conservation when simulating flexible boundaries. 
\end{abstract}
\section{Introduction}
The immersed methods are popular in modelling fluid-structure interaction (FSI) because they provide a computationally simple approach to deal with complex interfacial geometries and large deformations~\cite{peskin2002immersed, mittal2005immersed, griffith2020immersed}. For FSI problems involving thin interfaces, immersed methods typically use an Eulerian description of the fluid and a Lagrangian description of the interface. The immersed boundary (IB) method is one of the earliest of these types of methods. It was introduced by Peskin and uses integral transforms with Dirac delta function kernels to couple Eulerian and Lagrangian variables~\cite{peskin1972flow, peskin1977numerical, peskin2002immersed}. The immersed interface method (IIM) was developed by LeVeque and Li to sharply impose jump conditions in elliptic interface problems~\cite{leveque1994immersed}, and they extended the IIM to solve incompressible Stokes flow problems with flexible boundaries~\cite{leveque1997immersed}. Peskin and Printz~\cite{peskin1993improved} and Lai and Li~\cite{LAI2001149} established the jump conditions for the velocity and pressure in the incompressible Navier-Stokes equations in two and three spatial dimensions, respectively, and Li and Lai extended the IIM to treat the incompressible Navier-Stokes equations~\cite{li2001immersed}. Lee and LeVeque developed an alternative approach that combined the IB method and the IIM to model a flexible membrane~\cite{lee2003immersed}. To facilitate higher-order implementations of the IIM, Xu and Wang systematically derived both spatial and temporal jump conditions for the incompressible Navier-Stokes equations~\cite{xu2006systematic}, and applied the IIM to model rigid-body FSI~\cite{xu2006IIM}. Le~et~al. used the IIM to simulate multiple rigid and flexible structures~\cite{le2006immersed}. Meyer~et~al. combined the IIM and large eddy simulation to capture flow dynamics near fluid-structure interfaces under high Reynolds numbers and turbulent flows using a finite volume method~\cite{meyer2010conservative}. Kolahdouz~et~al. introduced a version of the IIM for general geometries described by $C^0$-continuous surface representations~\cite{kolahdouz2020immersed}.

Prior to this work, the immersed boundary-lattice Boltzmann method (IB-LBM) has been adopted to simulate fluid-structure interaction, including rigid body dynamics~\cite{feng2004immersed, jianhua2019FSI, niu2006momentum}, elastic filaments~\cite{tian2011efficient}, flexible sheets~\cite{zhu2011immersed} and large deformation of flexible beams~\cite{jianhua2019FSI}. The IB-LBM maintains the simplicity of the IB method, and the lattice Boltzmann method (LBM)~\cite{chen1998lattice} uses the linear lattice Boltzmann equation (LBE) to provide a nearly-incompressible fluid model that is simpler to parallelize than the incompressible Navier-Stokes equations because it permits a fully explicit time stepping scheme that does not require the solution of any global systems of equations. These previous IB-LBM are mostly based on so-called \textit{diffusive interface} IB methods, however, which effectively regularize discontinuities that generally are present at the immersed boundary. Kang and Hassan proposed a sharp interface IB-LBM by imposing the force density on exterior solid nodes nearest to the boundary~\cite{kang2011comparative}. However, the physical problems studied in their work are only related to static geometries, and it is not straightforward to generalize their approach to model flexible boundaries. This paper introduces an alternative approach to integrating \textit{sharp interface} IB approaches with LBM-based IB formulations, resulting in a methodology that is applicable to a wider range of problems. 

In this work, we first derive the jump conditions of the distribution function in the lattice Boltzmann equation that are related to applying a normal force along an immersed boundary. The jump conditions are then used to formulate an immersed interface-lattice Boltzmann method (II-LBM). The jump conditions arising from the normal component of the force are associated with pressure jump conditions in the conventional IIM for the incompressible Navier-Stokes equations. In this initial IIM-LBM formulation, the tangential forces are spread to the background grid using a standard diffuse interface IB approach. In addition, the velocity of the interface is determined using a standard IB interpolation scheme. The resulting methodology is similar to that of Lee and LeVeque~\cite{lee2003immersed}, except that here we use the LBM to model the fluid flow instead of the incompressible Navier-Stokes equations. Empirical results show that the II-LBM can achieve higher order of accuracy than the IB-LBM. 
Moreover, the comparison between the II-LBM and the IB-LBM using the simple external force term~\cite{he1997analytic} and Guo's external force term~\cite{guo2002discrete, guo2008analysis} shows that II-LBM has substantially superior volume conservation than the IB-LBM.

\section{Single-~and multi-relaxation-time IB-LBM }
This section introduces formulations of single-relaxation-time (SRT) and multi-relaxation-time (MRT) lattice Boltzmann methods and describes the integration of LBM formulations with conventional IB methods. Although the numerical tests considered in this paper are based on the more stable MRT model, the SRT model is included for completeness. Fig.~\ref{IIM_schematic} provides a schematic of a membrane immersed in a background fluid. Throughout this paper, $\Omega$ is the whole fluid domain, $\Gamma_t$ represents the immersed boundary at time $t$ (shown in Fig.~\ref{IIM_schematic}(a)), and the exterior and interior of the membrane at time $t$ are defined as $\Omega_t^+$ and $\Omega_t^-$, respectively, so that $\Omega = \Omega^+_t \cup \Omega^-_t$ (see Fig.~\ref{IIM_schematic}(b)).

\subsection{The immersed boundary-lattice Boltzmann method (IB-LBM)}
\subsubsection{Governing equations of the LBM}
Before going through the IB-LBM, the LBE governing the fluid flow is introduced. Here, the LBM based on both the SRT operator and the MRT operator are presented. 

\begin{figure}[htb!]
\sidesubfloat[]{\includegraphics[height=7cm]{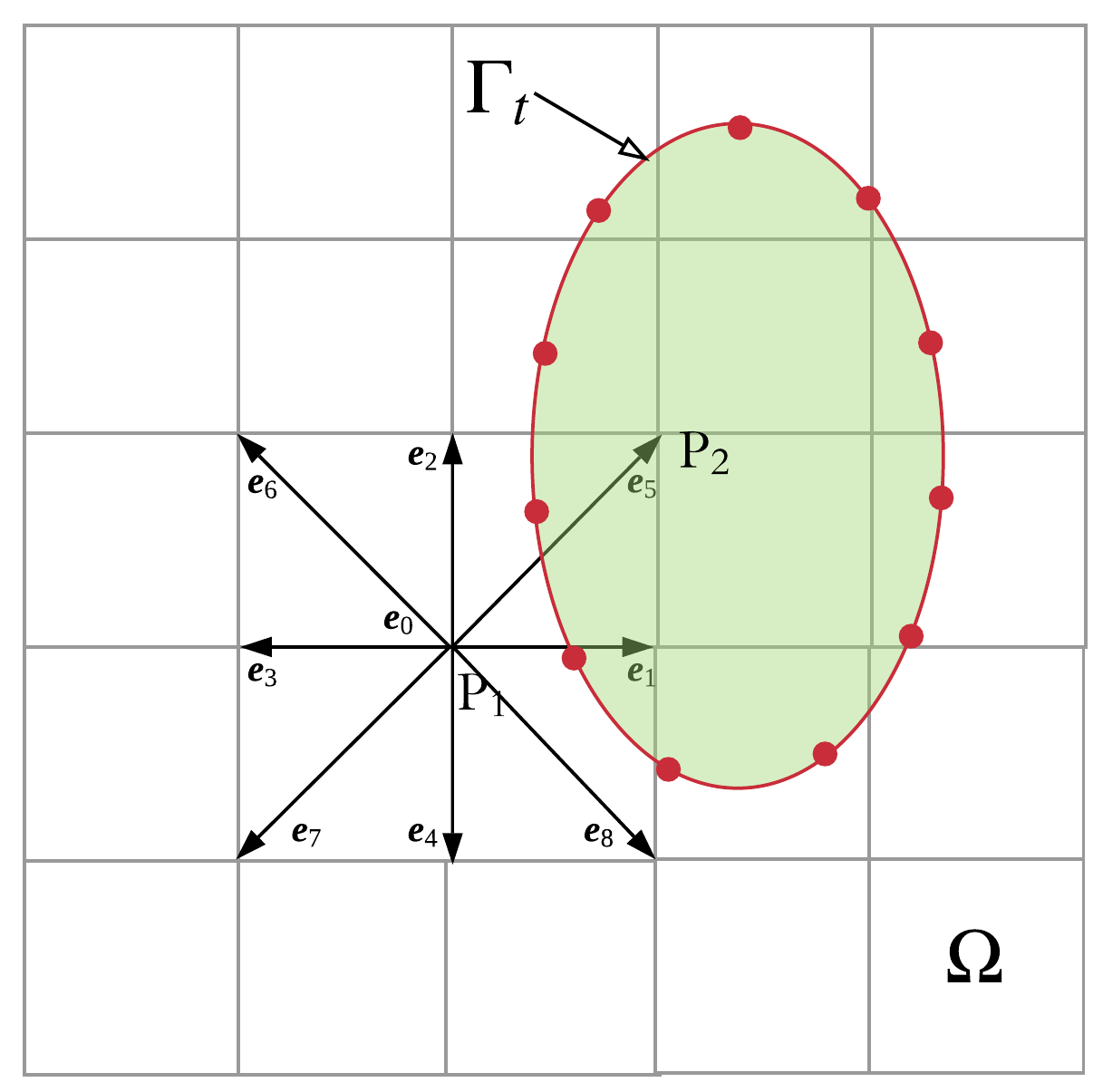}}
\sidesubfloat[]{\includegraphics[height=7cm]{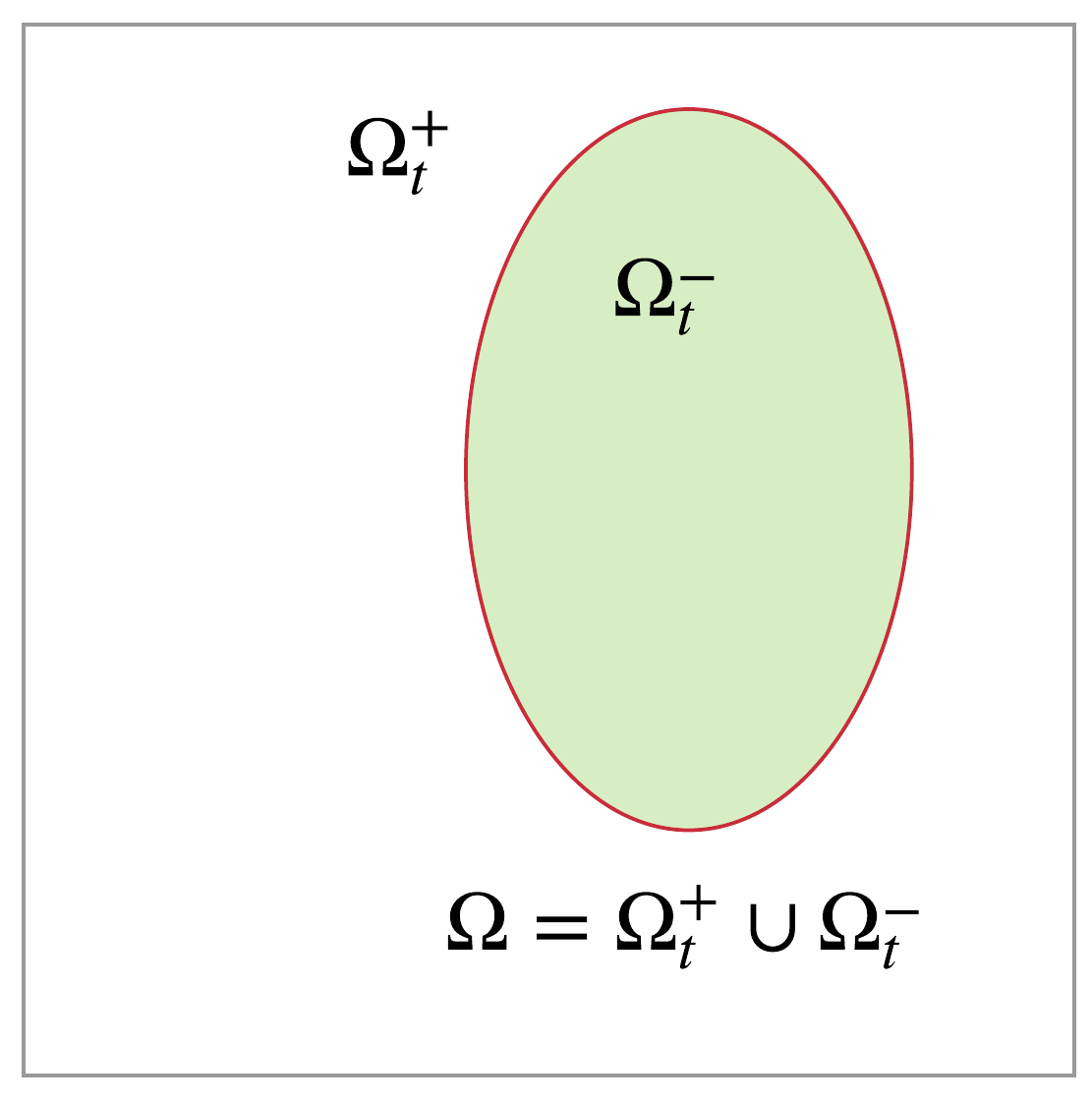}}
\caption{Schematic of a membrane ($\Gamma_t$) immersed in the fluid domain ($\Omega$). The interior fluid region ($\Omega_t^-$) is highlighted in green and the exterior fluid region ($\Omega_t^+$) is the remaining of the domain, which appears in white. $\bm{e}_i$ are velocity vectors for the D2Q9 model. }
\label{IIM_schematic}
\end{figure}

The equations for the distribution function $f_i$ of the SRT operator lattice Boltzman equation is
\begin{equation}
\frac{\partial f_i}{\partial t}+\textit{\textbf{e}}_i\cdot \nabla f_i=
-\frac{1}{\lambda}\left(f_i-f_i^{(\text{eq})}\right),\,i=0,1,2,...,Q,
\label{eq:LBM_srt_only}
\end{equation}
with $Q=18$ in the three-dimensional model (D3Q19), or by $Q=8$ in the two-dimensional model (D2Q9). Here, $\textbf{\textit{e}}_i$ are velocity vectors 
defined by $(0,0,0)$, $(\pm1,0,0)$, $(0,\pm1,0)$, $(0,0,\pm1)$, $(\pm1,\pm1,\pm1)$ for the D3Q19 model and $(0,0)$, $(\pm1,0)$, $(0,\pm1)$, $(\pm1,\pm1)$ for the D2Q9 model. 
Fig.~\ref{IIM_schematic} shows the $\bm{e}_i$ vectors associated with the identified point $\text{P}_1$ for the D2Q9 model. $\lambda$ is the relaxation parameter, and $f_i^{(\text{eq})}$ is the equilibrium distribution function.

\floatsetup[figure]{style=plain,subcapbesideposition=top}

By using the MRT operator in the lattice Boltzmann equation~\cite{lallemand2000theory, d2002multiple}, the evolution of the distribution functions becomes
\begin{equation}
\frac{\partial f_i}{\partial t}+\textit{\textbf{e}}_i\cdot \nabla f_i=-\textit{\textbf{M}}^{-1}\textit{\textbf{S}}_d\left(m_i(\textbf{\textit{x}},t)-m_i^{(\text{eq})}(\textbf{\textit{x}},t)\right),
\label{eq:LBM_mrt_only}
\end{equation}
in which $m_i$ represents the moments of the distribution function $f_i$. 
The matrix operator \textit{\textbf{M}} relates the distribution functions to the moments of the distribution functions via $\textit{\textbf{m}}=\textit{\textbf{M}}\textit{\textbf{f}}$, in which $\bm{m}=\{m_i\}$ and $\bm{f}=\{f_i\}$. 
Let $\bm{m}^{(\text{eq})}=\big\{m_i^{(\text{eq})}\big\}$ and $\bm{f}^{(\text{eq})}=\big\{f_i^{(\text{eq})}\big\}$ be the moments of equilibrium distribution functions and distribution functions, respectively. They are also related by using $\bm{M}$ as $\bm{m}^{(\text{eq})}=\bm{M}\bm{f}^{(\text{eq})}$. $\textit{\textbf{S}}_d=\text{diag}(s_0,s_1,s_2,...,s_{Q})$
is a diagonal matrix, the values of the elements in $\bm{S}_d$ will be given later. Although the SRT based LBM is extremely simple, the MRT based LBM is generally more robust~\cite{lallemand2000theory}.

\subsubsection{Governing equations of the IB-LBM}
The equations for the immersed boundary-lattice Boltzmann method (IB-LBM)~\cite{feng2004immersed, jianhua2018numerical} based on the simple external force term~\cite{he1997analytic} are
\begin{equation}
\frac{\partial f_i}{\partial t}+\textit{\textbf{e}}_i\cdot \nabla f_i=\text{RHS}+
\frac{1}{c_\text{s}^2}\omega_i\textbf{\textit{e}}_i\cdot \bm{g}(\textbf{\textit{x}},t),
\label{LBM_continuous}
\end{equation}
\begin{equation}
\bm{g}(\textbf{\textit{x}},t)=
\int_{\Gamma}\bm{G}(r,s,t)\,\delta(\textbf{\textit{x}}-\textbf{\textit{X}})
\,\text{d}r\text{d}s,
\label{eq:Lag_Eul_con_force}
\end{equation}
\begin{equation}
\frac{\partial \textbf{\textit{X}}(r,s,t)}{\partial t}=
\textbf{\textit{U}}(r,s,t)=
\int_{\Omega}\textit{\textbf{u}}(\textbf{\textit{x}},t)\delta (\textbf{\textit{x}}-\textbf{\textit{X}}(r,s,t))\,\text{d}\textbf{\textit{x}},
\label{eq:Lag_pos_update_con}
\end{equation}
\begin{equation}
\bm{G}(r,s,t)=\mathcal{G}(\bm{X},\bm{U},r,s,t).
\label{force_direct}
\end{equation}
Here, Eq.~(\ref{LBM_continuous}) accounts for the momentum of the fluid. The form of $\text{RHS}$ is determined by whether the SRT or MRT operator is used. For the SRT operator, $\text{RHS} = -\frac{1}{\lambda}\left(f_i-f_i^{(\text{eq})}\right)$, as in Eq.~(\ref{eq:LBM_srt_only}), and for the MRT operator, $\text{RHS} = -\textit{\textbf{M}}^{-1}\textit{\textbf{S}}_d\left(m_i(\textbf{\textit{x}},t)-m_i^{(\text{eq})}(\textbf{\textit{x}},t)\right)$, as in Eq.~(\ref{eq:LBM_mrt_only}). The physical configuration of the membrane $\Omega_t$ is taken as $\bm{X}(r,s,t)$, in which $(r,s)$ are curvilinear coordinates that parameterize the interface. The Lagrangian force density $\bm{G}(r,s,t)$ and the Eulerian force density $\bm{g}(\bm{x},t)$ are related to each other via Eq.~(\ref{eq:Lag_Eul_con_force}). $\delta(\textbf{\textit{x}}-\textbf{\textit{X}})$ is the Dirac delta function. In the numerical tests, we use Peskin's four-point regularized delta function~\cite{peskin2002immersed}. $\mathcal{G}$ is the functional that determines the interfacial force from the deformations and/or velocities of the immersed interface.

\subsection{Numerical approaches for the IB-LBM}
In the discrete equations, $\Delta x$ and $\Delta t$ are the grid spacing of the fluid and the time step size, respectively. We always use lattice units, so that $\Delta x=\Delta t=1$. The grid spacing between Lagrangian points in lattice units is defined as $\Delta X$.
\subsubsection{Discrete LBE}
The LBM is used to discretize the LBE for modelling the incompressible fluid flow. The relaxation parameter $\lambda$ is related to the kinematic viscosity $\nu$ by
\begin{equation}
\nu=c_{\text{s}}^2\left(\frac{\lambda}{\Delta t}-\frac{1}{2}\right)\Delta t,
\end{equation}
in which $c_\text{s}=\frac{1}{\sqrt{3}}$ is the speed of sound in lattice units, and $\omega_i$ is the weighting function defined by
\begin{equation}
\omega_i=
\begin{cases}
\frac{1}{3},&i=0,\\
\frac{1}{18},&1\leq i \leq6,\\
\frac{1}{36},&\text{otherwise},
\end{cases}
\end{equation}
for the D3Q19 model, or by
\begin{equation}
\omega_i=
\begin{cases}
\frac{4}{9},&i=0,\\
\frac{1}{9},&1\leq i \leq4,\\
\frac{1}{36},&\text{otherwise},
\end{cases}
\end{equation}
for the D2Q9 model.

For an incompressible fluid, the modified equilibrium distribution function~\cite{he1997lattice} is:
\begin{equation}
f_i^{(\text{eq})}=\omega_i\left[\frac{1}{c_\text{s}^2}p+\frac{\textit{\textbf{e}}_i\textit{\textbf{u}}}{c_\text{s}^2}
+\frac{(\textit{\textbf{e}}_i\textit{\textbf{u}})^2}{2c_\text{s}^4}-\frac{\textit{\textbf{u}}^2}{2c_\text{s}^2}\right].
\end{equation}
The macroscopic pressure $p$ and velocity $\textit{\textbf{u}}$ are obtained by
\begin{equation}
p=c_\text{s}^2\sum_{i=1}^{Q}f_i
\label{eq:LBM_pressure}
\end{equation}
and
\begin{equation}
\bm{u}=\sum_{i=1}^{Q}\bm{e}_i f_i.
\label{eq:LBM_velocity}
\end{equation}
The gradient of the distribution function ($\nabla f_i$) is discretized via $\text{Grad}(f_i)$ and is defined by
\begin{equation}
\text{Grad}(f_i)=\frac{f_i(\bm{x},t)-f_i(\bm{x}-\Delta \bm{x}_i,t)}{\Delta\bm{x}_i},
\label{eq:discretized_gradient}
\end{equation}
in which $\Delta\bm{x}_i=\bm{e}_i\Delta t$ is the grid spacing in the direction of $\bm{e}_i$. The discretized lattice Boltzmann equation with SRT becomes
\begin{equation}
\begin{aligned}
\frac{f_i(\bm{x},t+\Delta t)-f_i(\bm{x},t)}{\Delta t}+
\bm{e}_i\cdot 
\frac{f_i(\bm{x},t)-
f_i(\bm{x}-\Delta \bm{x}_i,t)}{\Delta\bm{x}_i}=
-\frac{1}{\lambda}\left(f_i-f_i^{(\text{eq})}\right).
\end{aligned}
\label{eq:dis_SRT_LBM}
\end{equation}
For the D3Q19 MRT model~\cite{d2002multiple}, $s_0=s_3=s_5=s_7=0,\, s_4=s_6=s_8,\, s_9=s_{11},\, s_{10}=s_{12},\, s_{13}=s_{14}=s_{15},\, s_{16}=s_{17}=s_{18}$. The nonzero elements in the diagonal matrices are $s_1=1.19\,\Delta t,\, s_2=s_{10}=1.4\,\Delta t,\, s_4=1.2\,\Delta t,\, s_9=s_{13}=\frac{\Delta t}{\lambda}$, and $s_{16}=1.98\,\Delta t$. In the D2Q9 MRT model~\cite{lallemand2000theory}, $s_0=s_3=s_5=0$, $s_1=1.63\,\Delta t$, $s_2=1.14\,\Delta t$, $s_4=s_6=1.92\,\Delta t$, and $s_7=s_8=\frac{\Delta t}{\lambda}$. Then the discretized lattice Boltzmann equation with MRT becomes
\begin{equation}
\begin{aligned}
\frac{f_i(\bm{x},t+\Delta t)-f_i(\bm{x},t)}{\Delta t}+
\bm{e}_i\cdot 
\frac{f_i(\bm{x},t)-
f_i(\bm{x}-\Delta \bm{x}_i,t)}{\Delta\bm{x}_i}=
-\textit{\textbf{M}}^{-1}\textit{\textbf{S}}_d\left(m_i(\textbf{\textit{x}},t)-m_i^{(\text{eq})}(\textbf{\textit{x}},t)\right).
\end{aligned}
\end{equation}

\subsubsection{Discrete IB-LBM}
The lattice Boltzmann equation with the simple external force term~\cite{he1997analytic} (i.e. Eq.~(\ref{LBM_continuous})) is discretized by
\begin{equation}
\begin{aligned}
\frac{f_i(\bm{x},t+\Delta t)-f_i(\bm{x},t)}{\Delta t}+
\bm{e}_i\cdot 
\frac{f_i(\bm{x},t)-
f_i(\bm{x}-\Delta \bm{x}_i,t)}{\Delta\bm{x}_i}=
\text{RHS}+\frac{1}{c_\text{s}^2}\omega_i\bm{e}_i\cdot
\bm{g}(\bm{x},t),
\end{aligned}
\label{eq:LBM_discrete_simple_force}
\end{equation}
in which the Eulerian force is determined via
\begin{equation}
\bm{g}(\bm{x},t)=\sum_{k}\bm{G}_k\,\delta_h(\bm{x}-\bm{X}_k)\,\Delta X.
\end{equation}
Here, $\delta_h(\bm{x}-\bm{X})$ is a regularized delta function. 

In the direct forcing IB method~\cite{feng2004immersed, jianhua2019FSI}, the Lagrangian force in Eq.~(\ref{force_direct}) is determined by
\begin{equation}
\bm{G}_k(t)=
\eta\left(\bm{U}_k^\text{D}(t)-\bm{U}_k^*(t)\right),
\label{Direct_forcing_force}
\end{equation}
in which $\bm{G}_k(t)$, $\bm{U}_k^\text{D}$, and $\bm{U}_k^*=
\sum_{\bm{x}}\bm{u}^*\,\delta_h(\bm{x}-\bm{X}_k)\,\Delta x^2
$ represent the force density, the desired Lagrangian velocity, and the interpolated Lagrangian velocity for the $k^\text{th}$th ($k = 1,2,\cdots,N$) Lagrangian point $\bm{X}_k$, respectively, and $\bm{u}^*$ is an intermediate Eulerian velocity without the effect the immersed boundary forcing. The penalty parameter $\eta$ is chosen as $\eta=\frac{1}{\Delta t}$.

In the classical IB method for a flexible membrane~\cite{peskin2002immersed}, the force on the $k^\text{th}$ Lagrangian node is calculated by
\begin{equation}
\bm{G}_k(t)=
\kappa \frac{\bm{X}_{k+1}(t)-2\bm{X}_{k}(t)
+\bm{X}_{k-1}(t)
}{\Delta X^2},
\label{Spring_forcing_force}
\end{equation}
in which $\kappa$ is a spring constant, and the dynamics of $\bm{X}_{k}$ are obtained by discretizing Eq.~(\ref{eq:Lag_pos_update_con}) via
\begin{equation}
\frac{\bm{X}_k(t+\Delta t)-\bm{X}_k(t)}{\Delta t}=\bm{U}_k(t)=
\sum_{\bm{x}}\bm{u}(\bm{x},t)\,\delta_h(\bm{x}-\bm{X})\,\Delta x^2.
\end{equation}

The IB-LBM based on Guo's external force terms~\cite{guo2002discrete, guo2008analysis} is used for some comparisons with the II-LBM introduced in this paper. Guo's external force terms have been demonstrated to recover the compressible Navier-Stokes equations more accurately. In their formulation, the last term in Eq.~(\ref{eq:LBM_discrete_simple_force}) is replaced by
\begin{equation}
G_i^\text{S}=\left(1-\frac{\Delta t}{2\lambda}\right)\omega_i\left(\frac{\bm{e}_i-\bm{u}}{c_\text{s}^2}+\frac{\bm{e}_i\cdot \bm{u}}{c_\text{s}^4}\cdot \bm{e}_i\right)\cdot\bm{g}\Delta t
\label{SRT_Guo_}
\end{equation} 
for SRT-LBM and by
\begin{equation}
\bm{G}^\text{M}=\bm{M}^{-1}\left(\bm{I}-\frac{1}{2}\bm{S}_d\right)\bm{M}\widetilde{\bm{G}}\Delta t
\label{MRT_Guo_}
\end{equation} 
for MRT-LBM, in which $\bm{G}^M=\left(G_0, G_1,..., G_q\right)^\text{T}$ and $\widetilde{\bm{G}}=\left(\widetilde{G}_0, \widetilde{G}_1,..., \widetilde{G}_q\right)^\text{T}$, with $\widetilde{G}_i=\omega_i\left(\frac{\bm{e}_i-\bm{u}}{c_\text{s}^2}+\frac{\bm{e}_i\cdot \bm{u}}{c_\text{s}^4}\cdot \bm{e}_i\right)\cdot\bm{g}$. Here, $\bm{I}$ is the identity matrix. In this case, the equilibrium function and velocity are
\begin{equation}
f_i^{(\text{eq})}=\omega_i\frac{p}{c_\text{s}^2}\left[1+\frac{\textit{\textbf{e}}_i\textit{\textbf{u}}}{c_\text{s}^2}
+\frac{(\textit{\textbf{e}}_i\textit{\textbf{u}})^2}{2c_\text{s}^4}-\frac{\textit{\textbf{u}}^2}{2c_\text{s}^2}\right]
\end{equation}
and
\begin{equation}
\frac{p}{c_\text{s}^2}\bm{u}=\sum_{i=1}^{Q}\textit{\textbf{e}}_i f_i+\frac{\Delta t}{2}\bm{g},
\end{equation}
in which $p$ is obtained by Eq.~(\ref{eq:LBM_pressure}).

\subsection{Rigid body dynamics}
For the numerical tests involving both the translation and rotation of rigid bodies in two spatial dimensions, the total force and torque acted on the structure are calculated by
\begin{equation}
\bm{F}^{\text{ext}}(t)=-\sum_{k=1}^N \bm{G}_k(t)\Delta X
+\rho_\text{f} V_{\text{s}}\frac{\bm{U}_\text{c}(r,s,t)-\bm{U}_\text{c}(r,s,t-\Delta t)}{\Delta t},
\label{Total_force}
\end{equation}
\begin{equation}
\begin{aligned}
T^{\text{ext}}(t)=&-\sum_{k=1}^N  \left(\bm{X}_k(r,s,t)-\bm{X}_{\text{c}}(r,s,t)\right)
\bm{\times}\bm{G}_k(t)\Delta X\\
&+\frac{\rho_\text{f}}{\rho_\text{s}} I_{\text{s}}\frac{W(r,s,t)-W(r,s,t-\Delta t)}{\Delta t},
\end{aligned}
\label{Total_torque}
\end{equation}
in which $\rho_\text{f}$ is the density of the fluid, $\rho_\text{s}$ is density of the structure, $V_\text{s}$ is the volume of the structure, $I_\text{s}$ is the moment of inertia, $\bm{X}_{\text{c}}(r,s,t)$, $\bm{U}_{\text{c}}(r,s,t)$, and $W(r,s,t)$ are the center of mass, velocity of the center of mass, and angular velocity of the structure at time $t$, respectively. The last terms in Eqs.~(\ref{Total_force}-\ref{Total_torque}) are to eliminate the effect of inertial forces from within the immersed interface~\cite{feng2009robust}.

After obtaining $\bm{F}^{\text{ext}}$ and $T^{\text{ext}}$, the acceleration and angular acceleration of the structure are determined by
\begin{equation}
\begin{aligned}
m\bm{a}(t+\Delta t)=\bm{F}^{\text{ext}}(t),\\
I_{\text{s}} w(t+\Delta t)=T^{\text{ext}}(t),
\end{aligned}
\end{equation}
in which $m$, $\bm{a}$, $I_{\text{s}}$, and $w$ are the mass, acceleration, moment of inertia, and angular acceleration of the rigid body, respectively. Then the velocity and angular velocity of the rigid body are updated via
\begin{equation}
\begin{aligned}
\bm{U}_{\text{c}}(t+\Delta t)=\bm{U}_{\text{c}}(t)+\Delta t\,\bm{a}(t+\Delta t),\\
W(t+\Delta t)=W(t)+\Delta t\, w(t+\Delta t).
\end{aligned}
\end{equation}
Ultimately, the new position and orientation angle of the structure are determined by
\begin{equation}
\begin{aligned}
\bm{X}_{\text{c}}(t+\Delta t)&=\bm{X}_{\text{c}}(t)+\bm{U}_{\text{c}}(t)\Delta t+\frac{1}{2}(\Delta t)^2(\bm{a}(t)+\bm{a}(t+\Delta t)),\\
\theta(t+\Delta t)&=\theta(t)+W(t)\Delta t+\frac{1}{2}(\Delta t)^2(w(t)+w(t+\Delta t)),
\end{aligned}
\end{equation}
in which $\theta(t)$ is the orientation angle of the rigid body at time $t$.

\section{The immersed interface-lattice Boltzmann method (II-LBM)}
The II-LBM depends on the jump condition across the fluid-solid interface. This section formulates the jump conditions in the framework of the LBE. The resulting II-LBM can be viewed as a correction of the IB-LBM considering the formulated jump conditions. 
\subsection{The derivation of the jump conditions}
For a scalar field $\psi$, the jump is defined as
\begin{equation}
\begin{aligned}
\llbracket \psi(\bm{X},t)\rrbracket
=& \lim_{\epsilon\,\downarrow\,0}\psi(\bm{X}+\epsilon\bm{n}(\bm{X},t),t)-
\lim_{\epsilon\,\downarrow\,0}\psi(\bm{X}-\epsilon
\bm{n}(\bm{X},t),t)
\\
=& \psi^+(\bm{x},t)-\psi^-(\bm{x},t),
\end{aligned}
\end{equation}
in which $\psi(\bm{X},t)$ is a position along the immersed interface, $\llbracket\cdot\rrbracket$ is the operator for the jump, and $\psi^+(\bm{x},t)$ and  $\psi^-(\bm{x},t)$ are limiting values of $\psi$ approaching interface position $\bm{X}$ from exterior region $\Omega_t^+$ and interior region $\Omega_t^-$, respectively. The unit normal vector at the interface is
\begin{equation}
\textbf{\textit{n}}=\frac{\frac{\partial \textbf{\textit{X}}}{\partial r}\times \frac{\partial \textbf{\textit{X}}}{\partial s}}{\left|\frac{\partial \textbf{\textit{X}}}{\partial r}\times \frac{\partial \textbf{\textit{X}}}{\partial s}\right|}.
\end{equation}
For rigid body simulations, $\left|\frac{\partial \textbf{\textit{X}}}{\partial r}\times \frac{\partial \textbf{\textit{X}}}{\partial s}\right| = 1$.

The jump conditions of the distribution function based on the simple external force term~\cite{he1997analytic} are
\begin{equation}
\llbracket f_i\rrbracket=
\frac{\omega_i}{c_\text{s}^2}\frac
{\bm{G}(r,s,t)\cdot \textbf{\textit{n}}}
{\left|\frac{\partial \textbf{\textit{X}}}{\partial r}\times \frac{\partial \textbf{\textit{X}}}{\partial s}\right|}.
\label{jump_f0to8}
\end{equation}
Following Lai and Li~\cite{LAI2001149}, we can derive the above jump conditions in three spatial dimensions. Let us choose a banded domain $\Omega_{\epsilon,t}$ enclosing the immersed membrane $\Gamma_t$ with outer and inner subregions $\Omega_{\epsilon,t}^+$ and $\Omega_{\epsilon,t}^-$. Here we denote $\epsilon$ as a small distance parameter. Suppose $\phi(\bm{x})$ is any smooth function with compact support on $\Omega_{\epsilon,t}$. Multiplying $\phi(\bm{x})$ on both sides of Eq.~(\ref{LBM_continuous}) and integrating over $\Omega_{\epsilon,t}$, we obtain 
\begin{equation}
\begin{aligned}
& \int_{\Omega_{\epsilon,t}}\frac{\partial f_i}{\partial t}\,\phi\,\text{d}\bm{x}+
 \int_{\Omega_{\epsilon,t}}\left(\bm{e}_i\cdot \nabla f_i\right)
\phi\,\text{d}\bm{x}\\
=\int_{\Omega_{\epsilon,t}}\text{RHS}\,\phi\,\text{d}\bm{x}
+ &
\frac{1}{c_\text{s}^2}\omega_i\bm{e}_i\cdot
\int_{\Omega_{\epsilon,t}}\left(\int_0^{L_r}\int_0^{L_s}
\bm{G}(r,s,t)\delta (\bm{x}-\bm{X}(r,s,t))
\,\text{d}r\text{d}s\right)\phi\,
\text{d}\bm{x},
\end{aligned}
\label{LBM_int}
\end{equation}
in which $L_r$ and $L_s$ are the limits of the parameterized coordinate systems. The first term and third term in the above equation go to zero as $\epsilon\rightarrow 0$ because $f_i$ and $f_i^{(\text{eq})}$ are continuous and bounded. Using the fact that $\Omega_{\epsilon,t}$ encloses $\Gamma_t$,  the last term in the above equation can be simplified, and the Eq.~(\ref{LBM_int}) becomes
\begin{equation}
\int_{\Omega_{\epsilon,t}}\left(\textit{\textbf{e}}_i\cdot \nabla f_i\right)
\phi\,\text{d}\bm{x}=
\frac{1}{c_\text{s}^2}\omega_i\textbf{\textit{e}}_i\cdot
\int_0^{L_r}\int_0^{L_s}
\bm{G}(r,s,t)\,\phi\,
\text{d}r\text{d}s.
\label{simple_continuous}
\end{equation}
By the divergence theorem, the first term in Eq.~(\ref{simple_continuous}) is
\begin{equation}
\textit{\textbf{e}}_i\int_{\Omega_{\epsilon,t}}\left(\nabla f_i\right)\phi\,\text{d}\bm{x}=
\textbf{\textit{e}}_i
\left(
\int_{\Omega_{\epsilon,t}^+} f_i\bm{n}\phi\,\text{d}a+
\int_{\Omega_{\epsilon,t}^-} f_i(-\bm{n})\phi\,\text{d}a-
\int_{\Omega_{\epsilon,t}}f_i\,\nabla\phi\, \text{d}\textbf{\textit{x}}
\right).
\label{Eq_above}
\end{equation}
in which $\text{d}a=\left|\frac{\partial \textbf{\textit{X}}}{\partial r}\times \frac{\partial \textbf{\textit{X}}}{\partial s}\right|\text{d}r\text{d}s$ is the area of the surface element. The last term in Eq.~(\ref{Eq_above}) converges to zero as $\epsilon\rightarrow 0$ because $f_i$ is bounded.
Then Eq.~(\ref{Eq_above}) becomes
\begin{equation}
\bm{e}_i\int_{\Omega_{\epsilon,t}}\left(\nabla f_i\right)\phi\,\text{d}\bm{x}=
\bm{e}_i\int_0^{L_r}\int_0^{L_s}\llbracket f_i\rrbracket 
\textbf{\textit{n}} \phi
\left|\frac{\partial \textbf{\textit{X}}}{\partial r}\times \frac{\partial \textbf{\textit{X}}}{\partial s}\right|
\text{d}r\text{d}s.
\label{Eq_jump_int}
\end{equation}
Combining Eq.~(\ref{simple_continuous}) and Eq.~(\ref{Eq_jump_int}) and taking advantage of the fact that $\phi$ is arbitrary, we obtain
\begin{equation}
\llbracket f_i\rrbracket\textbf{\textit{n}}\left|\frac{\partial \textbf{\textit{X}}}{\partial r}\times \frac{\partial \textbf{\textit{X}}}{\partial s}\right|=\frac{\omega_i}{c_\text{s}^2}\bm{G}(r,s,t), \quad i\neq 0.
\end{equation}
Since $\textbf{\textit{n}}$ is the unit normal vector,
\begin{equation}
\llbracket f_i\rrbracket\left|\frac{\partial \textbf{\textit{X}}}{\partial r}\times \frac{\partial \textbf{\textit{X}}}{\partial s}\right|=\frac{\omega_i}{c_\text{s}^2}
\textbf{\textit{n}}\cdot
\bm{G}(r,s,t), \quad i\neq 0.
\label{JUMP_f1to8}
\end{equation}
Then the only unknown jump condition for the distribution function is
$\llbracket f_0\rrbracket$. Apply the jump operator to both sides of Eq.~(\ref{eq:LBM_pressure}):
\begin{equation}
\begin{aligned}
& \llbracket p\rrbracket =\left\llbracket c_\text{s}^2\sum_if_i\right\rrbracket=c_\text{s}^2\sum_i\llbracket f_i\rrbracket\\
& \quad\;\;=
\frac
{\bm{G}(r,s,t)\cdot \textbf{\textit{n}}}
{\left|\frac{\partial \textbf{\textit{X}}}{\partial r}\times \frac{\partial \textbf{\textit{X}}}{\partial s}\right|}
\sum_{i\neq 0}\omega_i+c_\text{s}^2\llbracket f_0\rrbracket\\
& \quad\;\;=
\frac
{\bm{G}(r,s,t)\cdot \textbf{\textit{n}}}
{\left|\frac{\partial \textbf{\textit{X}}}{\partial r}\times \frac{\partial \textbf{\textit{X}}}{\partial s}\right|}-
\frac
{\bm{G}(r,s,t)\cdot \bm{n}}
{\left|\frac{\partial \bm{X}}{\partial r}\times \frac{\partial \bm{X}}{\partial s}\right|}\,\omega_0+c_\text{s}^2\llbracket f_0\rrbracket.
\end{aligned}
\end{equation}
Because Eq.~(\ref{JUMP_f1to8}) only deals with the normal part of the boundary forces, the equations for the jump conditions of the distribution function should correspond to the jump condition of the pressure~\cite{LAI2001149},
\begin{equation}
\llbracket p\rrbracket=\frac
{\bm{G}(r,s,t)\cdot \bm{n}}
{\left|\frac{\partial \bm{X}}{\partial r}\times \frac{\partial \bm{X}}{\partial s}\right|}.
\end{equation}
From the above two equations, the jump condition $\llbracket f_0 \rrbracket$ is
\begin{equation}
\llbracket f_0\rrbracket=
\frac{\omega_0}{c_\text{s}^2}\frac
{\bm{G}(r,s,t)\cdot \bm{n}}
{\left|\frac{\partial \bm{X}}{\partial r}\times \frac{\partial \bm{X}}{\partial s}\right|}.
\label{JUMP_f0}
\end{equation}
Combining Eq.~(\ref{JUMP_f1to8}) and Eq.~(\ref{JUMP_f0}) gives the jump condition for the distribution functions in Eq.~(\ref{jump_f0to8}). The velocity jump can be determined via
\begin{equation}
\llbracket\bm{u}\rrbracket=\left\llbracket \sum_if_i\bm{e}_i\right\rrbracket=\sum_i\bm{e}_i\llbracket f_i\rrbracket=
\sum_i\bm{e}_i
\frac
{\omega_i \bm{G}(r,s,t)\cdot \textbf{\textit{n}}}
{c_\text{s}^2\left|\frac{\partial \bm{X}}{\partial r}\times \frac{\partial \bm{X}}{\partial s}\right|}.
\end{equation}
After rearrangement, we have
\begin{equation}
\llbracket \bm{u}\rrbracket=
\frac{\bm{G}(r,s,t)\cdot\bm{n}}
{c_\text{s}^2\left|\frac{\partial \bm{X}}{\partial r}\times \frac{\partial \bm{X}}{\partial s}\right|}
\sum_i\bm{e}_i\omega_i.
\label{no_slip}
\end{equation}
Because the weighting function $\omega_i$ is the same for opposite directions of $\bm{e}_i$, we have $\Sigma_i\bm{e}_i\omega_i=\bm{0}$. This allows us to reduce Eq.~(\ref{no_slip}) to $\llbracket \bm{u}\rrbracket=\bm{0}$, which is the continuity of the velocity field across the membrane.

\subsection{Correction of the IB-LBM considering the jump conditions}
Jump conditions are imposed in the discretized SRT-LBM by including appropriate correction terms in Eq.~(\ref{eq:LBM_discrete_simple_force}). Because we consider only the normal component of the force in determining the jump conditions, this is similar to the application of the pressure jump condition on the gradient of the pressure when adopting the finite difference method to solve the incompressible Navier-Stokes equations~\cite{kolahdouz2020immersed}. The discretized gradient of the distribution function ($\text{Grad}(f_i)$ in Eq.~(\ref{eq:discretized_gradient})) can be written considering the jump conditions as
\begin{equation}
\text{Grad}(f_i)=\frac{f_i(\bm{x},t)-f_i(\bm{x}-\Delta \bm{x}_i,t)}{\Delta\bm{x}_i}-
\frac{\llbracket f_i(\bm{x},t)\rrbracket}{\Delta\bm{x}_i}.
\end{equation}
Therefore, the discretized lattice Boltzmann equation with the simple external force term~\cite{he1997analytic} (Eq.~(\ref{eq:LBM_discrete_simple_force})) should be corrected by
\begin{equation}
\begin{aligned}
& \frac{f_i(\bm{x},t+\Delta t)-f_i(\bm{x},t)}{\Delta t}+
\bm{e}_i\cdot 
\left(
\frac{f_i(\bm{x},t)-
f_i(\bm{x}-\Delta \bm{x}_i,t)}{\Delta\bm{x}_i}-
\frac{\llbracket f_i(\bm{x},t)\rrbracket}{\Delta\bm{x}_i}
\right)\\=
\quad &
\text{RHS}+\frac{1}{c_\text{s}^2}\omega_i\bm{e}_i\cdot
\bm{g}^{||}(\bm{x},t),
\end{aligned}
\label{eq:discretized_LBE_He}
\end{equation}
in which $\bm{g}^{||}(\bm{x},t)$ is force of the immersed boundary node in the tangential direction. Because the jump conditions determined here for the distribution function only impose the pressure jump condition associated with the normal component of the interfacial force. In the present method, the tangential force is treated via a discretized integral transform with a regularized delta function kernel, as in the conventional IB and IB-LB methods. The external force density on the fluid node is obtained by spreading the Lagrangian force on the IB node
\begin{equation}
\bm{g}^{||}(\bm{x},t)=
\sum_k\bm{G}^{||}_k(t)\delta
(\bm{x}-\bm{X})\Delta X,
\label{Eq_IBM_tangential}
\end{equation}
in which $\bm{G}^{||}_k=\bm{G}_k-
\bm{G}_k\cdot\bm{n}$ is the tangential part of the Lagrangian force.
Eq.~(\ref{eq:discretized_LBE_He}) can be rewritten as
\begin{equation}
\begin{aligned}
f_i(\bm{x},t+\Delta t)=
f_i(\bm{x}-\Delta \bm{x}_i,t)+\text{RHS}\Delta t+
\llbracket f_i(\bm{x},t+\Delta t)\rrbracket
+\frac{1}{c_\text{s}^2}\omega_i\bm{e}_i\cdot
\bm{g}^{||}(\bm{x},t+\Delta t)\Delta t.
\end{aligned}
\label{LBM_WITH_JUMP}
\end{equation}

In those cases where Guo's external force terms~\cite{guo2002discrete, guo2008analysis} are used, the Eulerian force $\bm{g}$ used to calculate $G_i^\text{S}$ in Eq.~(\ref{SRT_Guo_}) and $\tilde{G}_i$ in Eq.~(\ref{MRT_Guo_}) should be replaced by $\bm{g}^{||}$. For the point P$_1$ in Fig.~\ref{IIM_schematic}, only the distribution functions $f_3$ and $f_7$ are updated by Eq.~(\ref{LBM_WITH_JUMP}) with $\llbracket f_i(\bm{x},t)\rrbracket\neq 0$. For the point P$_2$ in Fig.~\ref{IIM_schematic}, only the distributions functions $f_1$, $f_5$ and $f_8$ are updated considering the jump conditions.

To implement the jump conditions of the distribution functions on a certain fluid node in Eq.~(\ref{jump_f0to8}), the corresponding Lagrangian force $\bm{G}(r,s,t)$ on the structure surface, and the normal direction of the fluid node to the structure surface have to be determined. In this paper, the level set method~\cite{li2006immersed} is used when simulating rigid body simulations, and a cubic spline representation is adopted in simulating flexible boundaries~\cite{lee2003immersed}.

\section{Results}
This section provides results from several benchmark problems. For the problems studied in this paper, Guo's external force terms are only used in Sec.~4.2 and the multi-relaxation-time operator LBM is used for all the cases.
\subsection{Uniform normal force along a circle}
\begin{figure}[htb!]
  \includegraphics[height=6cm]{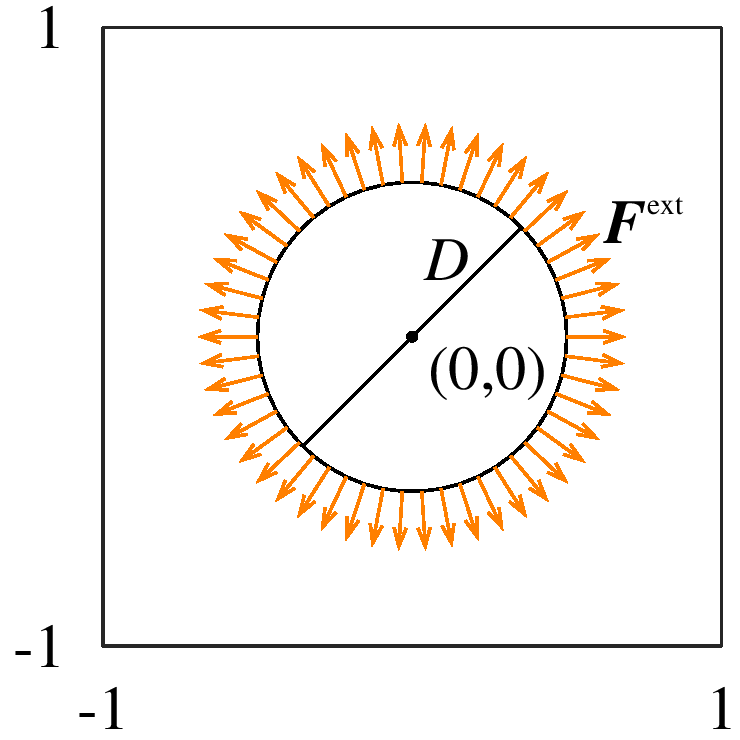}
  \caption{Schematic of uniformly distributed normal force along a circular interface.}
\label{fig:Singular_schematic}
\end{figure}

We first consider a circular cylinder immersed in the fluid with evenly distributed force along the normal direction of the surface. This case is the same as the one considered by Li and Lai~\cite{li2001immersed}. The schematic of the present calculation is shown in Fig.~\ref{fig:Singular_schematic}. The diameter of the cylinder is $D$. The fluid domain is $[-D,D]\times [-D,D]$ with the cylinder located at its center. The boundaries of the outer domain are considered as outflow and handled by the non-equilibrium extrapolation strategy~\cite{zhao2002non}. The Reynolds number and the viscosity are chosen as $Re=100$ and $\nu=0.05$. The characteristic velocity is $U=\frac{\nu\,Re}{D}$, and the magnitude of the normal interfacial forces is chosen to be $\frac{\bm{F}^\text{ext}}{\rho_\text{f}\,U^2}=1$. By varying $U$ and $D$, different cases are obtained. The normalized $L_2$-norm and $L_\infty$-norm are defined by
\begin{equation}
\begin{aligned}
L_2\text{-norm}=\frac{\sqrt{\left(\sum\limits_{\bm{x}} 
(u_x-u_x^\text{e})^2+(u_y-u_y^\text{e})^2\right)/N_\text{f}}
}{U},\\
L_\infty\text{-norm}=\frac{\max\limits_{\bm{x}}\,\max(|u_x-u_x^\text{e}|,\,|u_y-u_y^\text{e}|)
}{U},
\end{aligned}
\end{equation}
in which $N_\text{f}$ is the number of fluid points of the domain $\Omega$ and $u_x$ and $u_y$ are fluid velocities in the $x$ and $y$ directions, respectively. The superscript ``e" indicates the exact velocity in the fluid domain. In this case, the exact velocity is $u_x=u_y=0$.

\begin{figure}[htb!]
  \sidesubfloat[]{\includegraphics[height=7cm]{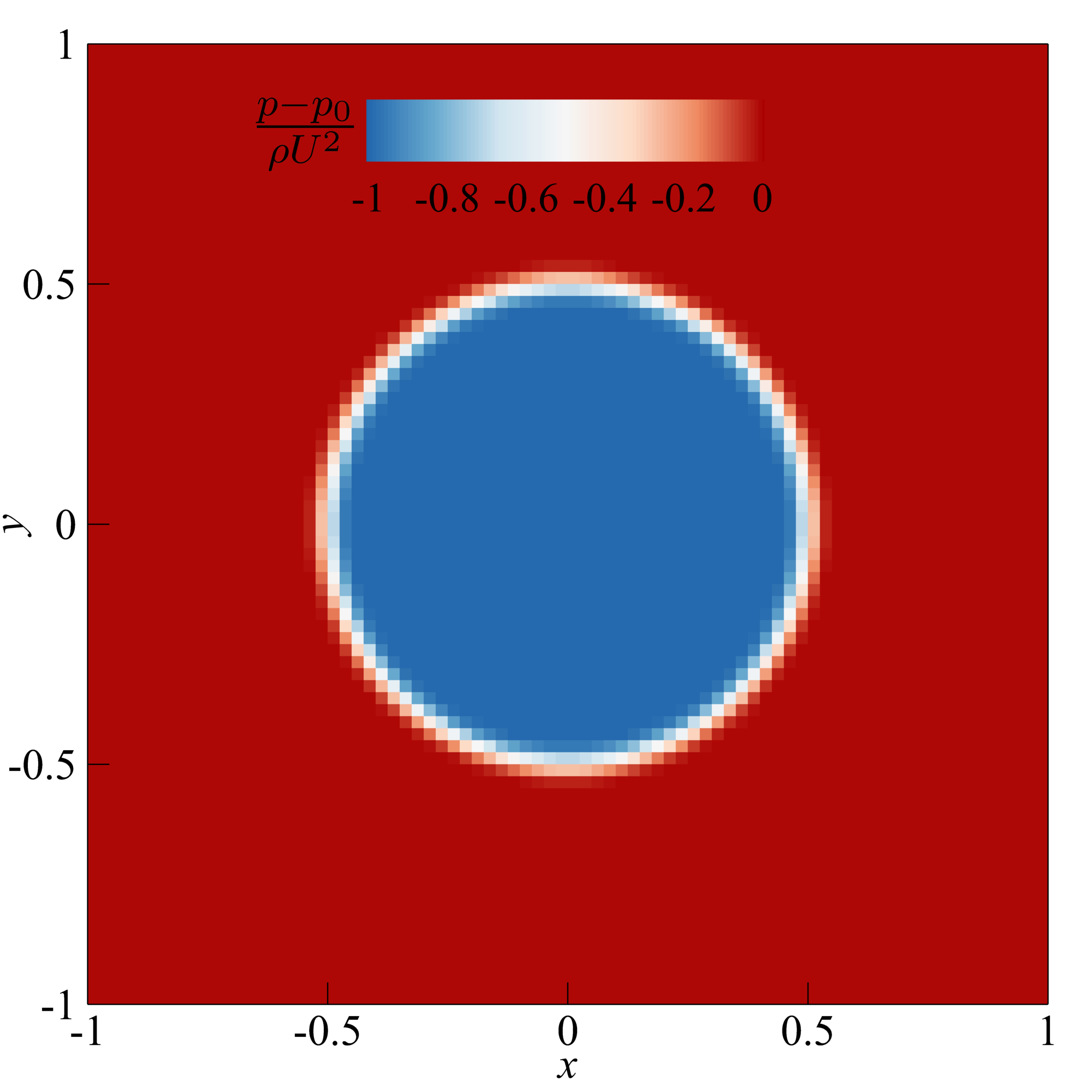}}
  \sidesubfloat[]{\includegraphics[height=7cm]{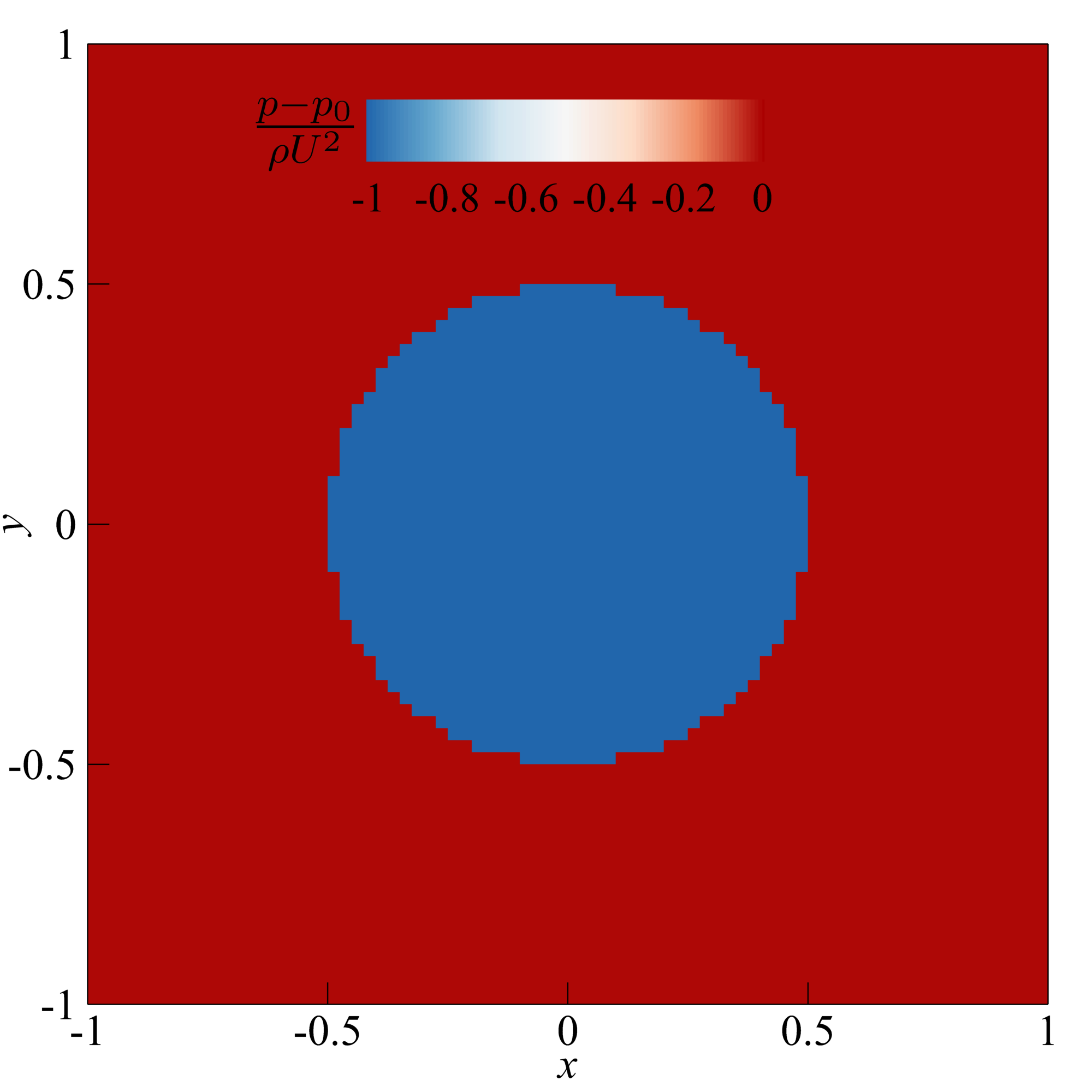}}
  \caption{Pressure fields for evenly distributed normal force along a circle calculated by using (a) the direct forcing IB-LBM and (b) the II-LBM with $D=40\Delta x$ at $t^\text{end}=D/U$.}
\label{Pressure_singular}
\end{figure}

\begin{figure}[htb!]
  \sidesubfloat[]{\includegraphics[height=7cm]{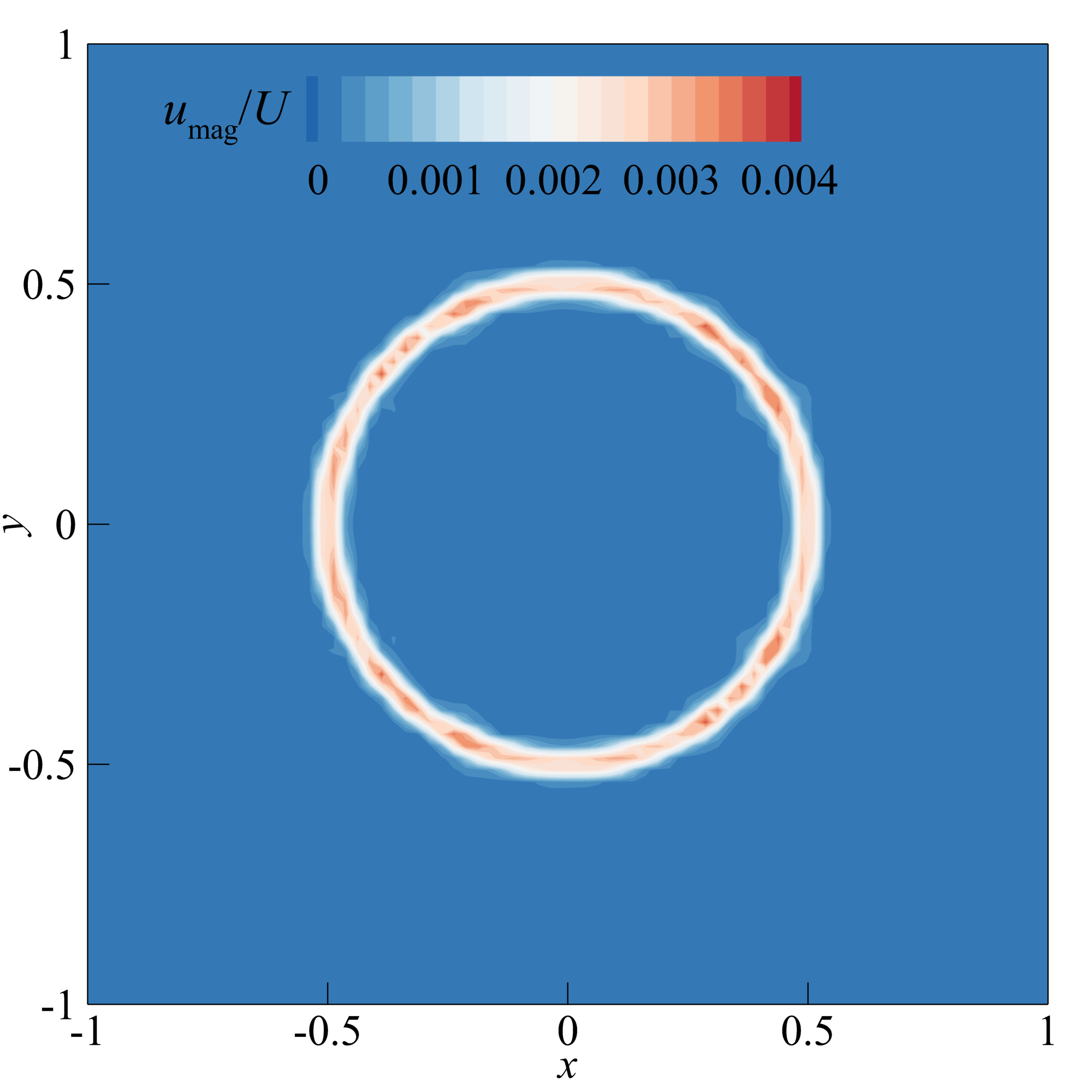}}
  \sidesubfloat[]{\includegraphics[height=7cm]{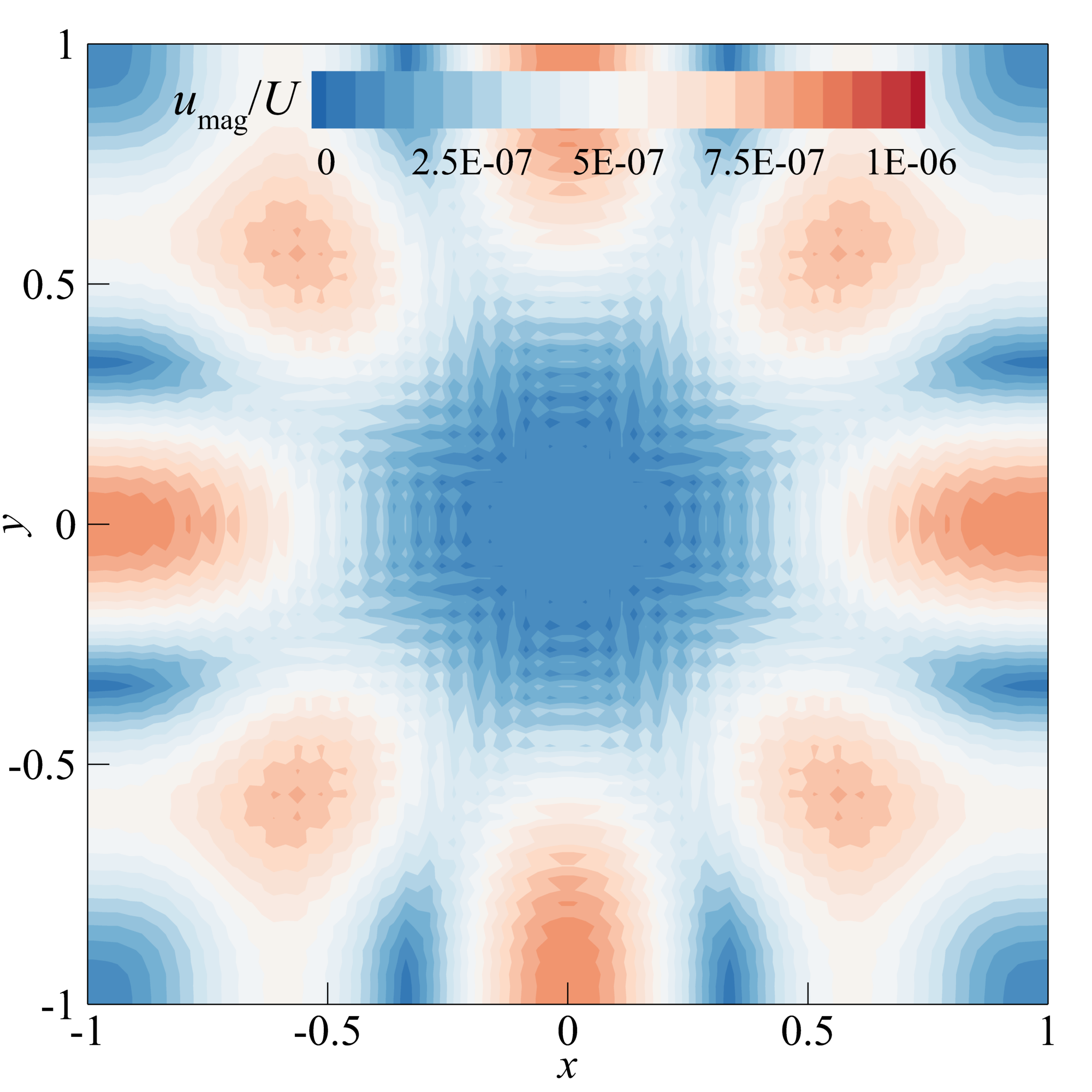}}
  \caption{Velocity magnitude fields for evenly distributed normal force along a circle calculated by using (a) the direct forcing IB-LBM and (b) the II-LBM with $D=40\Delta x$ at $t^\text{end}=D/U$.}
\label{Velocity_singular}
\end{figure}

Errors are assessed at time $t^\text{end}=D/U$. The results of the error and the order of accuracy calculated by the IIM and the direct forcing IB method in this case are shown in Table \ref{tab:Uniform_force}. The pressure and velocity fields calculated by using the IB method and the IIM at $t^\text{end}=D/U$ are presented in Fig.~\ref{Pressure_singular} and \ref{Velocity_singular}, respectively. The pressure fields clearly show that the IIM gives piecewise constant pressure whereas the IB method regularizes the pressure discontinuity. The velocity magnitude obtained by the IIM in Fig.~\ref{Velocity_singular}(b) is much smaller than that in Fig.~\ref{Velocity_singular}(a) obtained by the IB method. This is because the jump conditions of the distribution function balance the normal forces while satisfying the continuity of the velocity field across the interface in Eq.~(\ref{no_slip}). On the other hand, the IB method corrects the velocity fields to balance to boundary forces.
  
\begin{table}[htb!]
    \centering
    \caption{Comparison of $L_2$-norm and $L_\infty$-norm for uniform distributed force along the surface of a circular cylinder by the direct forcing IB method and the IIM in different meshes.}
    \begin{tabular}{p{1cm}p{1.5cm}p{1.0cm}p{1.5cm}p{1.0cm}p{0.05cm}p{1.5cm}p{1.0cm}p{1.5cm}p{1.0cm}}
      \toprule
      \multirow{2}{*}[-2pt]{$D$} 
 & \multicolumn{4}{c}{$L_2$-norm} & & \multicolumn{4}{c}{$L_\infty$-norm} \\
 \cmidrule{2-5} \cmidrule{7-10}
 & IB-LBM & Order & II-LBM & Order & & IB-LBM & Order & II-LBM & Order\\
	  \midrule
$20\Delta x$ & 1.45e-3 & & 1.86e-6 & & & 6.39e-3 & & 3.31e-6 & \\
$40\Delta x$ & 5.30e-4 & 1.37 & 4.62e-7 & 2.01 && 3.23e-3 & 0.99 & 8.46e-7 & 1.96\\
$80\Delta x$ & 1.90e-4 & 1.39 & 1.14e-7 & 2.03 & & 1.70e-3 & 0.95 & 2.12e-7 & 1.99\\
$160\Delta x$ & 6.77e-5 & 1.40 & 2.84e-8 & 2.01  & & 8.34e-4 & 1.02 & 5.32e-8 & 1.99\\
      \bottomrule
    \end{tabular}
    \label{tab:Uniform_force}
  \end{table}
 
\subsection{Dynamic motion of a thin elastic interface}
For the incompressible Navier-Stokes equations, using a staggered-grid spatial discretization~\cite{griffith2012volume} or specialized differencing operators \cite{peskin1993improved} can greatly reduce spurious volume (area) change resulting from the non-divergence free velocity field of the Lagrangian points by the IB method. However, the lattice Boltzmann equation is generally solved with an explicit time integration scheme using a collocated grid. To our knowledge, a staggered-grid LBM has not been developed. Lee and LeVeque~\cite{lee2003immersed} found that the immersed interface method with the pressure jump condition has better volume conservation than the IB method. To validate that the II-LBM proposed in this paper also preserves the volume for flexible bodies, the dynamic motion of a thin elastic interface is studied.

\begin{figure}[htb!]
  \includegraphics[height=6cm]{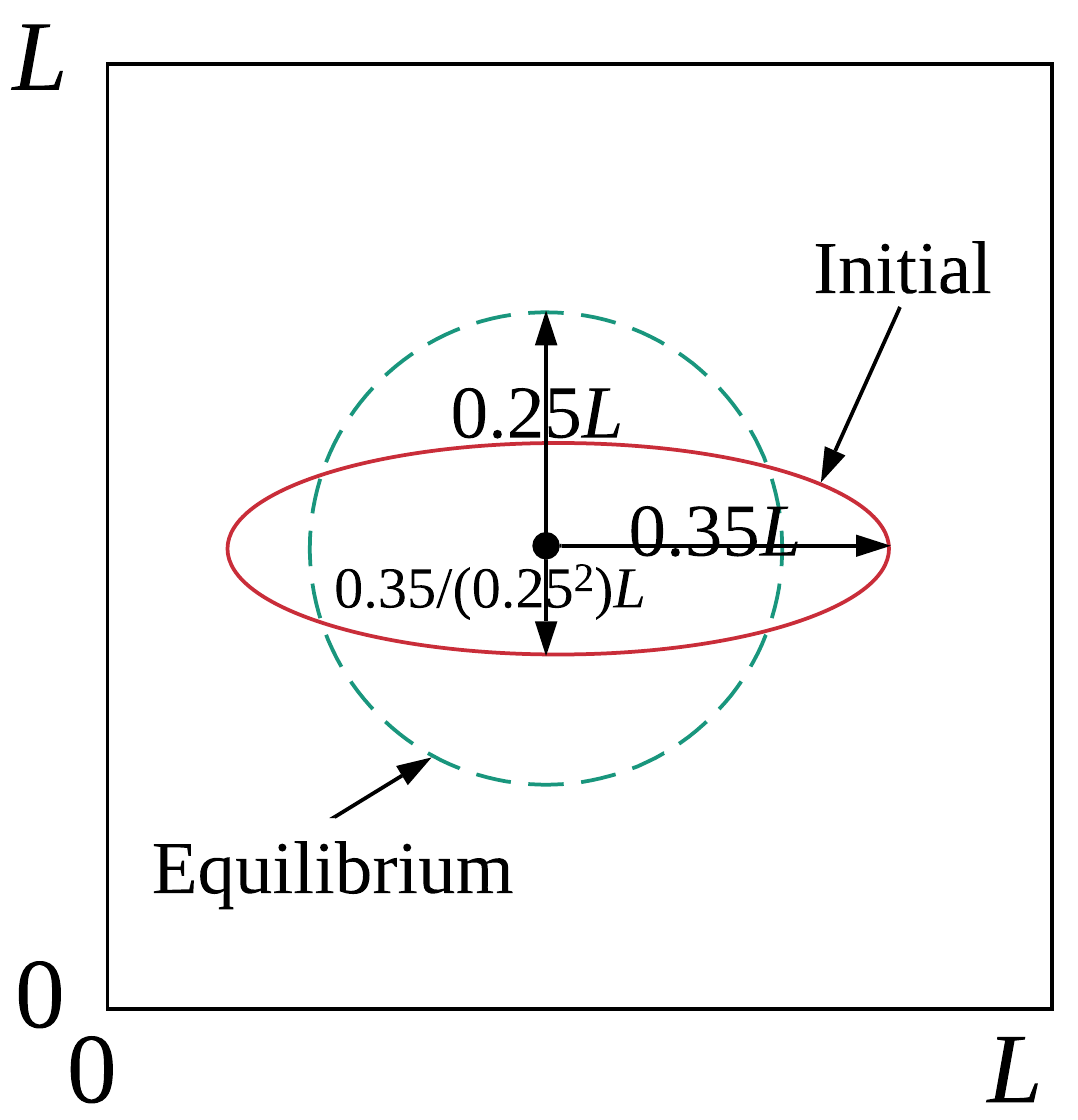}
  \caption{Schematic of dynamic motion of a thin elastic interface.}
\label{fig:IIM_schematic}
\end{figure}

\begin{figure}[htb!]
  \includegraphics[height=7cm]{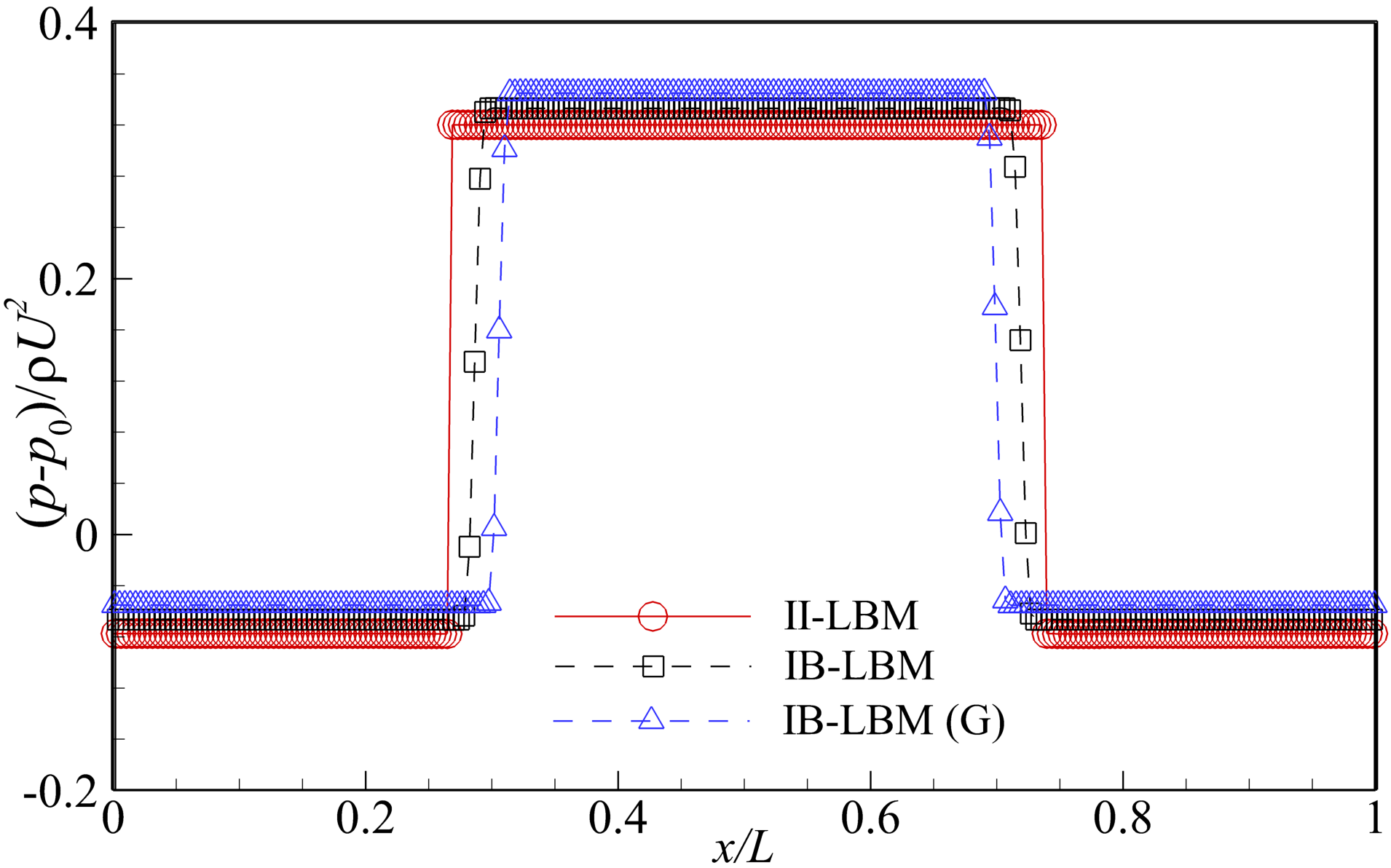}
  \caption{Pressure distribution along $y/L=0.5$ for dynamic motion of a thin elastic interface for the II-LBM, the IB-LBM using the simple external force term~\cite{he1997analytic} and the IB-LBM (G) using Guo's external force term~\cite{guo2008analysis}.}
\label{IBM_FSI}
\end{figure}

\begin{figure}[htb!]
  \subfloat[]{\includegraphics[height=6cm]{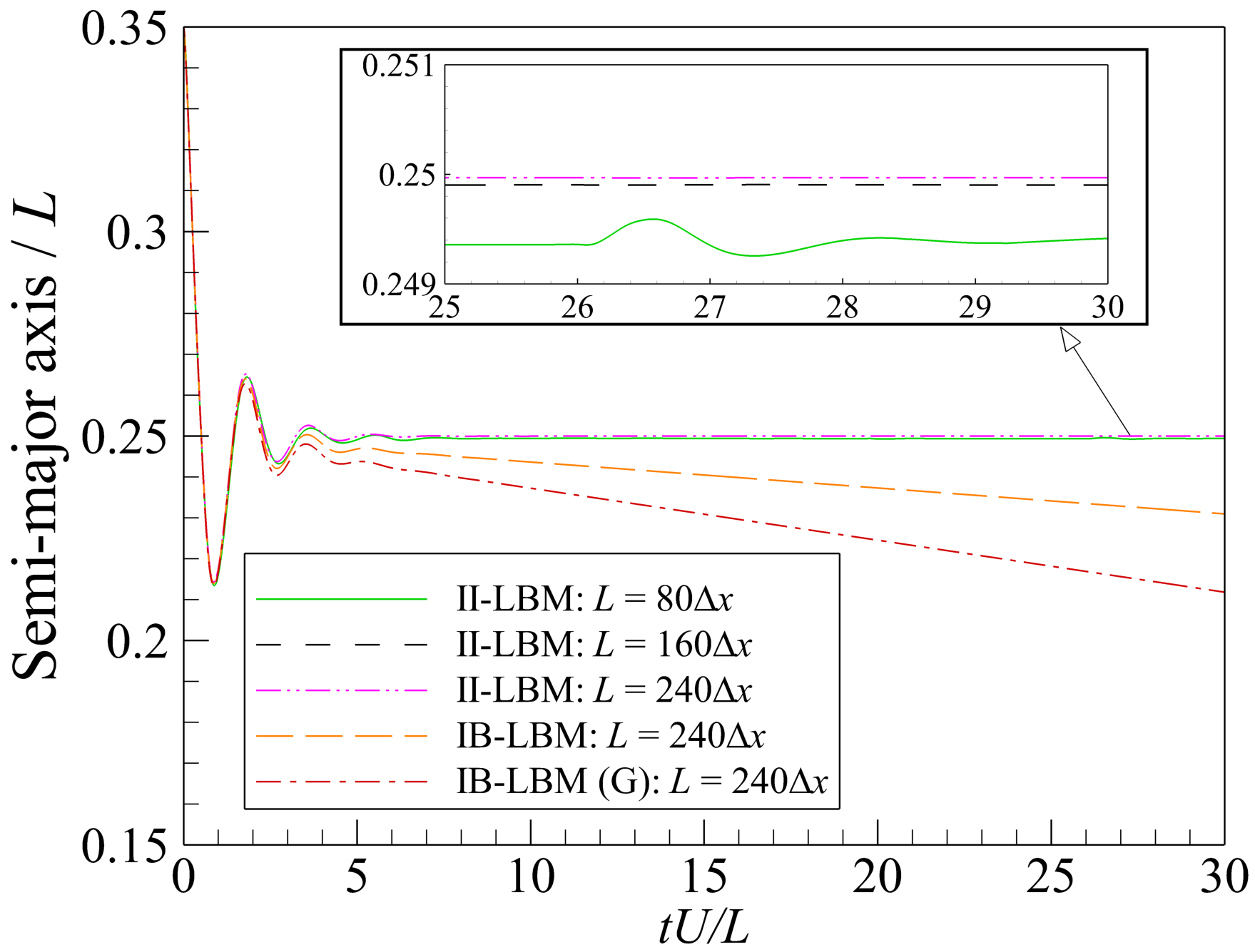}}
  \subfloat[]{\includegraphics[height=6cm]{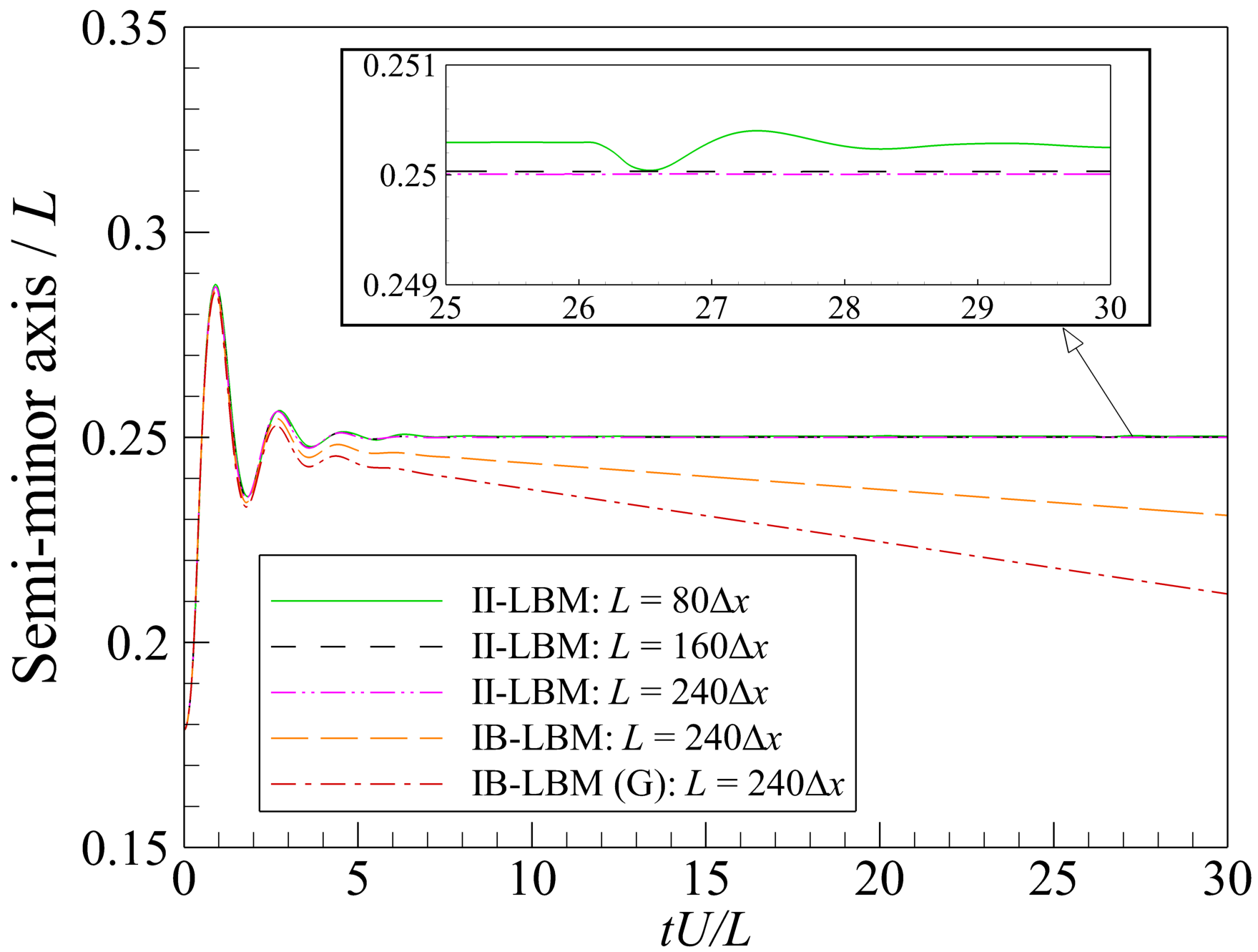}}\\
  \caption{Evolution of (a) semi-major axis and (b) semi-minor axis for the dynamic motion of a thin elastic ellipse interface calculated by the II-LBM, the IB-LBM with the simple external force term and the IB-LBM with Guo's external force term~\cite{guo2008analysis}.}
\label{Ellipse_extension}
\end{figure}

The schematic of the computational domain is shown in Fig.~\ref{fig:IIM_schematic}, where the initial shape is an ellipse interface and the equilibrium shape is expected to be a circular interface. A square box of side length $L$ is chosen as the computational domain. The boundaries of the outer domain are set to no-slip walls by using the non-equilibrium extrapolation strategy~\cite{zhao2002non}. The ellipse interface is initially positioned at $(0.5D,0.5D)$ with semi-major and semi-minor axes of $a = 0.35L$ and $b = \frac{0.25^2}{0.35}L$, respectively. The nondimensional spring constant and viscosity are set to $\overline{\kappa}=\frac{\kappa}{U^2L}=0.1$ and $\nu=0.02$, respectively. The resulting spring constant $\kappa$ will be used in Eq.~(\ref{Spring_forcing_force}) to calculate the spring force. The Reynolds number is set to $Re=100$, and the characteristic velocity is determined as $U=\frac{\nu\,Re}{L}$. The distance between two adjacent Lagrangian points are taken as $\Delta X=\frac{1}{3}\Delta x$ for the IB-LBM and $\Delta X=2\Delta x$ for the II-LBM with respect to the circular interface in its equilibrium state of radius $0.25L$. Three different meshes, corresponding to $L=80\Delta x, 160\Delta x$, and $240\Delta x$, are used. The first period of oscillation ends at approximately $tU/L=0.55\frac{3.0625^{\log_{10} (\overline{\kappa})}}{\overline{\kappa}}$, which is the same as that indicated by Griffith~\cite{griffith2012volume}. As can be seen from the pressure distribution at $tU/D=30$ at $y/L=0.5$ calculated by $L=240\Delta x$ in Fig.~\ref{IBM_FSI}, the II-LBM gives a sharp jump at the interface of $x=0.25L$ whereas the other two methods produce a regularized pressure discontinuity. Fig.~\ref{Ellipse_extension} shows the changes of the length of the semi-major and semi-minor axes against three different approaches. As expected, both the IB-LBM with multi-relaxation-time using a simple external force term~\cite{he1997analytic} and Guo's external force term~\cite{guo2008analysis} result in substantial volume loss. However, the volume of the ellipse barely changes in time by using the II-LBM, and the final result of $L=240\Delta x$ gives a 0.01\% error of the semi axis as compared to the analytical result. As seen in Fig.~\ref{Ellipse_extension}, although the II-LBM shows some oscillations for the coarsest grid, the oscillations converge rapidly and become unnoticeable for finer grids. To quantify the three different approaches adopted in the performance of preserving volume, Table~\ref{tab:area_loss} is presented to show the area loss per dimensionless time $\overline{\Delta V}_\text{loss}$ and total area loss $\overline{V^\text{total}_\text{loss}}$ at $t^\text{end}=30L/U$
\begin{equation}
\overline{\Delta V}_\text{loss}=\frac{\Delta V_\text{loss}}{LU\Delta t},
\end{equation}
and
\begin{equation}
\overline{V^\text{total}_\text{loss}}=\frac{V^\text{total}_{\text{loss}}}{L^2},
\end{equation}
in which $\Delta V_\text{loss}$ and $V^\text{total}_{\text{loss}}$ are dimensional volume of loss per unit time and total volume loss at time $t$.

{\renewcommand{\arraystretch}{1.5}  
\begin{table}[htb!]
    \centering
    \caption{Area loss per unit time and total area loss at $t^\text{end}=30L/U$ obtained by the IB-LBM with the simple external force term~\cite{he1997analytic}, the IB-LBM with Guo's external force term~\cite{guo2008analysis}, and the II-LBM.}
    \begin{tabular}{p{1.2cm}p{1.05cm}p{1.05cm}p{1.05cm}p{0.05cm}p{1.05cm}p{1.05cm}p{1.05cm}p{0.05cm}p{1.05cm}p{1.05cm}p{1.05cm}}
      \toprule
      \multirow{2}{*}[-2pt]{} 
  	  & \multicolumn{3}{c}{IB-LBM}& &\multicolumn{3}{c}{IB-LBM (G)}& &\multicolumn{3}{c}{II-LBM}\\
	  \cmidrule{2-4}\cmidrule{6-8}\cmidrule{10-12}
	  & $80\Delta x$ & $160\Delta x$ & $240\Delta x$& &  $80\Delta x$ & $160\Delta x$ & $240\Delta x$& & $80\Delta x$ & $160\Delta x$ & $240\Delta x$ \\
	  \midrule
$\overline{\Delta V}_\text{loss}$ &$2.42\text{e}{-3}$ & $1.42\text{e}{-3}$ & $8.82\text{e}{-4}$& &$3.70\text{e}{-3}$& $2.42\text{e}{-3}$& $1.78\text{e}{-3}$ & & $7.87\text{e}{-6}$  & $1.83\text{e}{-6}$ & $6.28\text{e}{-7}$  \\
$\overline{V^\text{total}_\text{loss}}$  & $7.93\text{e}{-2}$ & $4.29\text{e}{-2}$ & $2.87\text{e}{-2}$& &$1.38\text{e}{-1}$& $7.93\text{e}{-2}$& $5.51\text{e}{-2}$& & $2.64\text{e}{-4}$  & $5.29\text{e}{-5}$ & $2.35\text{e}{-5}$ \\
      \bottomrule
    \end{tabular}
    \label{tab:area_loss}
  \end{table}
} 

\subsection{Uniform flow around a circular cylinder}
Flow around a circular cylinder is frequently used as a benchmark to check the accuracy of a numerical algorithm. The left and right boundaries are taken as inflow and outflow, respectively. The inflow velocity is $U$.~Freestream boundary conditions are used on the side walls.~The computational domain is $[0,40D]\times[0,20D]$ with the circular cylinder of diameter $D$ located at $(10D,10D)$. The Strouhal number is defined as
\begin{equation}
\text{St}=\frac{fD}{U},
\label{Strouhal}
\end{equation}
in which $f$ is the frequency of the vortex shedding. The drag and lift coefficients are defined as
\begin{equation}
C_\text{D}=\frac{2F^{\text{ext}}_x}{\rho_\text{f}\,U^2D},
\label{Drag_co}
\end{equation}
\begin{equation}
C_\text{L}=\frac{2F^{\text{ext}}_y}{\rho_\text{f}\,U^2D},
\label{Lift_co}
\end{equation}
in which $\rho_\text{f}$ is the density of the fluid. $F^{\text{ext}}_x$ and $F^{\text{ext}}_y$ are external forces in the $x$ and $y$ directions, respectively.

\begin{figure}[htb!]
  \sidesubfloat[]{\includegraphics[height=5cm]{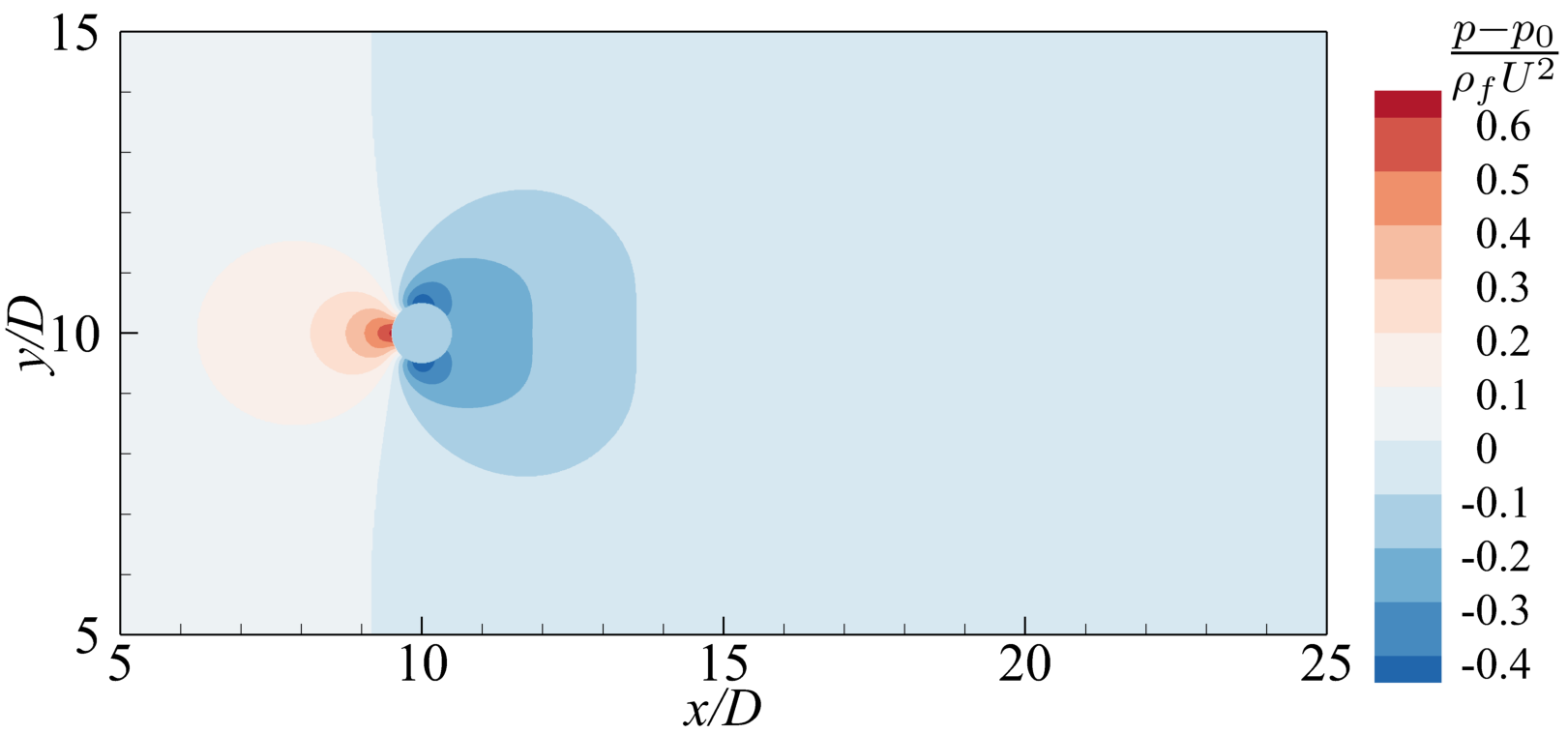}}\\
  \sidesubfloat[]{\includegraphics[height=5cm]{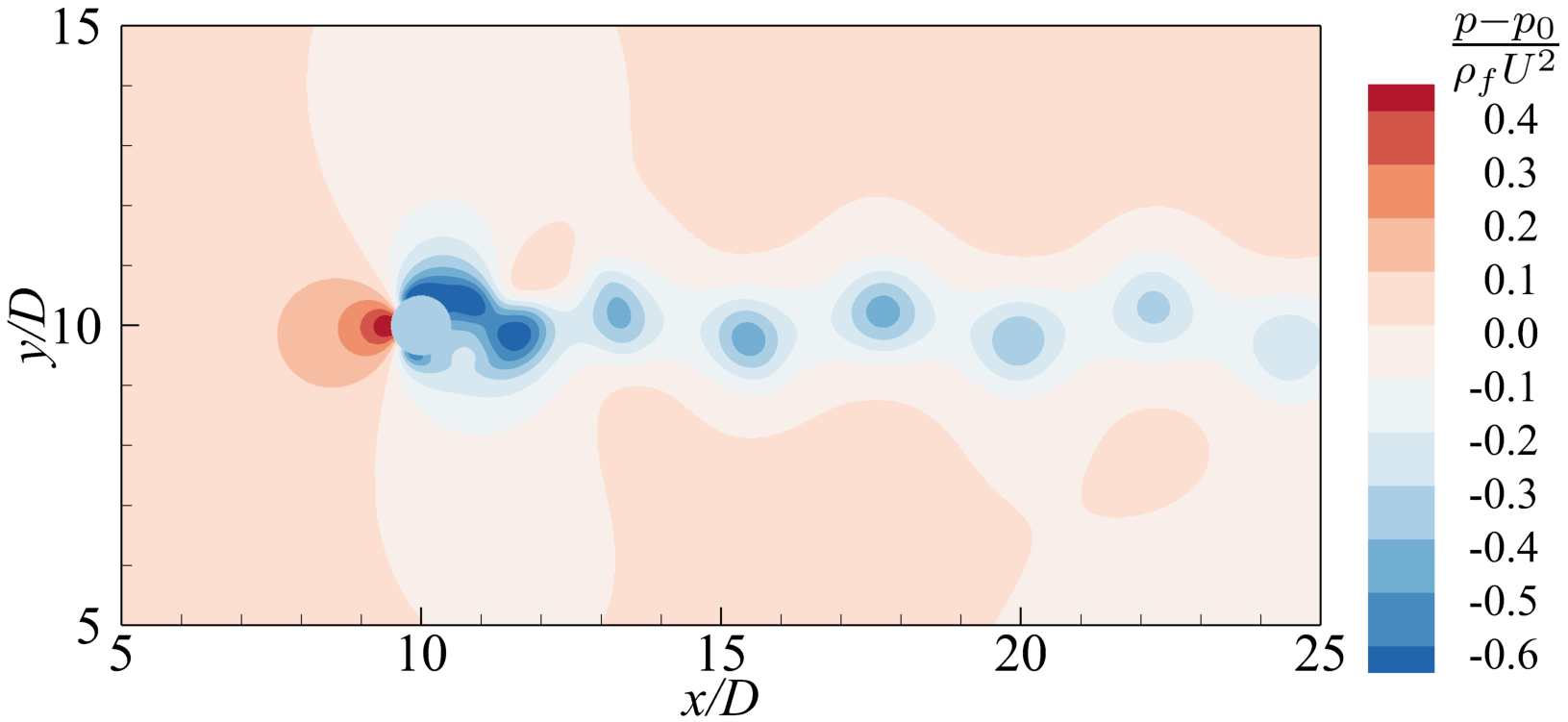}}
  \caption{Pressure fields for flow around a circle cylinder calculated by the II-LBM using at (a) $Re=40$ and (b) $Re=200$ with $D=80\Delta x$.}
\label{fig:Pressure_cylinder_static}
\end{figure}

\begin{figure}[htb!]
  \sidesubfloat[]{\includegraphics[height=7.5cm]{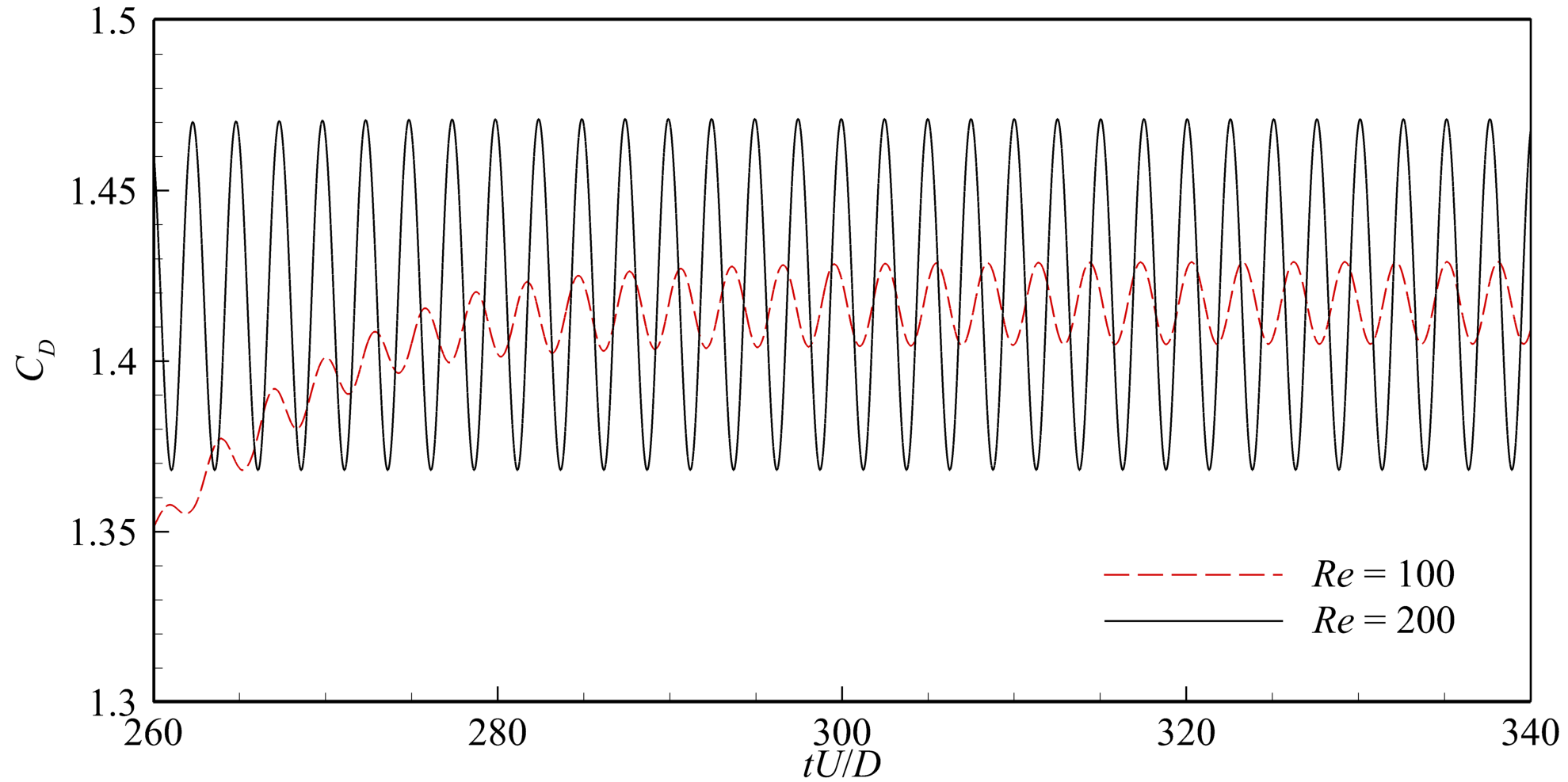}}\\
  \sidesubfloat[]{\includegraphics[height=7.5cm]{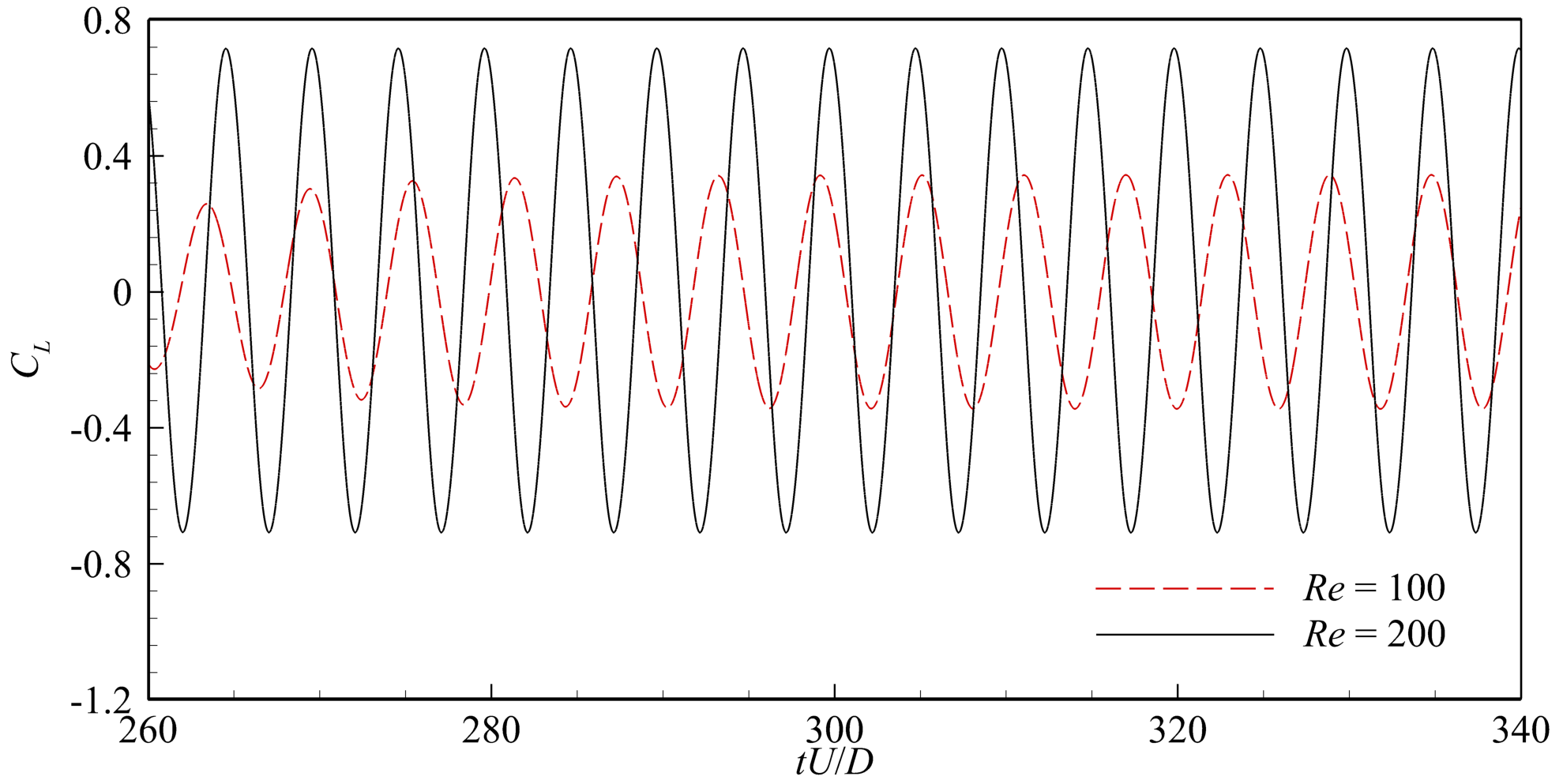}}
  \caption{(a) Drag coefficients and (b) lift coefficients for flow around a circular cylinder at $Re=100$ and 200 calculated by the II-LBM with $D=80\Delta x$.}
\label{fig:Drag_lift_Re100_200}
\end{figure}

Table.~\ref{tab:Flow_cylinder_Re20_40} lists the drag coefficient and recirculation length ($\text{R}_\text{l}$) and angle of separation ($\Lambda$) for the steady flow cases of $Re = 20$ and 40. As the Reynolds number increases, the flow changes from steady flow ($Re \leq 47$) to unsteady flow and the Von K{\'a}rm{\'a}n vortex street appears. To quantify the accuracy of the II-LBM in identifying the flow characteristics of unsteady flow, a comparison of average drag coefficient ($\overline{C}_\text{D}$), lift coefficient ($C_\text{L}$) and Strouhal number (St) is presented in Table.~\ref{tab:Flow_cylinder_Re100_200}. The pressure fields in Fig.~\ref{fig:Pressure_cylinder_static} show that the II-LBM can capture the discontinuities in the pressure field at the interface. Fig.~\ref{fig:Drag_lift_Re100_200} present the time history of the drag and lift coefficients at $Re=100$ and 200. These figures indicate that the periodic vortex shedding is successfully simulated by the II-LBM.

\begin{table}[htb!]
    \centering
    \caption{Flow around a circular cylinder calculated by the II-LBM for different meshes at $Re=20$ and $Re=40$.}
    \begin{tabular}{p{4.5cm}p{1.5cm}p{1.5cm}p{1.5cm}p{1.5cm}p{1.5cm}p{1.5cm}}
      \toprule
      \multirow{2}{*}[-2pt]{References} 
  	  & \multicolumn{3}{c}{$Re=20$ ($\nu=0.1$)} &\multicolumn{3}{c}{$Re=40$ ($\nu=0.05$)}\\
	  \cmidrule{2-4}\cmidrule{5-7}
	  & $C_\text{D}$  & $\text{R}_\text{l}/D$ & $\Lambda$ (deg.) & $C_\text{D}$  & $\text{R}_\text{l}/D$ & $\Lambda $(deg.) \\
	  \midrule
Tritton~\cite{tritton1959experiments} (Exp.) & 2.22 & --- & --- & 1.48 &  --- & --- \\
Calhoun~\cite{calhoun2002cartesian} (Num.) & 2.19 & 0.91 & 45.5 & 1.62 & 2.18 & 54.2 \\
Xu and Wang~\cite{xu2006IIM} (Num.) & 2.23 & 0.92 & 44.2 & 1.66 & 2.21 & 53.5 \\
Kolahdouz et al.~\cite{kolahdouz2020immersed} & 2.10 & 0.93 & 44.4 & 1.58 & 2.31 & 54.1\\
II-LBM ($D=20\Delta x$) & 2.220  & 1.000 & 42.4 & 1.663  & 2.481 & 50.1\\
II-LBM ($D=40\Delta x$) & 2.195  & 0.968 & 43.4 & 1.642  & 2.380 & 52.6\\
II-LBM ($D=80\Delta x$) & 2.189  & 0.953 & 44.3 & 1.632  & 2.321 & 53.1\\
      \bottomrule
    \end{tabular}
    \label{tab:Flow_cylinder_Re20_40}
  \end{table}
  
\begin{table}[htb!]
    \centering
    \caption{Flow around a circular cylinder calculated by the II-LBM for different meshes at $Re=100$ and $Re=200$.}
    \begin{tabular}{p{4.5cm}p{1.5cm}p{1.5cm}p{1.5cm}p{1.5cm}p{1.5cm}p{1.5cm}}
      \toprule
      \multirow{2}{*}[-2pt]{References} 
  	  & \multicolumn{3}{c}{$Re=100$ ($\nu=0.02$)} &\multicolumn{3}{c}{$Re=200$ ($\nu=0.01$)}\\
	  \cmidrule{2-4}\cmidrule{5-7}
 & $\overline{C}_\text{D}$ & $C_\text{L}$ & St & $\overline{C}_\text{D}$ & $C_\text{L}$ & St\\
	  \midrule
Roshko~\cite{roshko1954development} (Exp.) & --- & --- & 0.167 & --- & --- & 0.190 \\
Williamson~\cite{williamson1988defining} (Exp.) & --- & --- & 0.166 & --- & --- & 0.197 \\
Calhoun~\cite{calhoun2002cartesian} (Num.) & 1.33 & $\pm$0.298 & 0.175 & 1.17 & $\pm$0,67 & 0.202\\
Xu and Wang~\cite{xu2006IIM} (Num.) & 1.423 & $\pm$0.340 & 0.171 & 1.42 & $\pm$0.66 & 0.202 \\
II-LBM ($D=20\Delta x$) & 1.458 & $\pm$0.357 & 0.163 & 1.471 & $\pm$0.654 & 0.193\\
II-LBM ($D=40\Delta x$) & 1.436 & $\pm$0.344 & 0.167 & 1.442 & $\pm$0.718 & 0.197\\
II-LBM ($D=80\Delta x$) & 1.417 & $\pm$0.344 & 0.169 & 1.419 & $\pm$0.713 & 0.199\\
      \bottomrule
    \end{tabular}
    \label{tab:Flow_cylinder_Re100_200}
  \end{table}
  
\floatsetup[figure]{style=plain,subcapbesideposition=top}
  
\subsection{Prescribed translation of a circular cylinder in a resting background fluid}
\begin{figure}[htb!]
  \includegraphics[height=5cm]{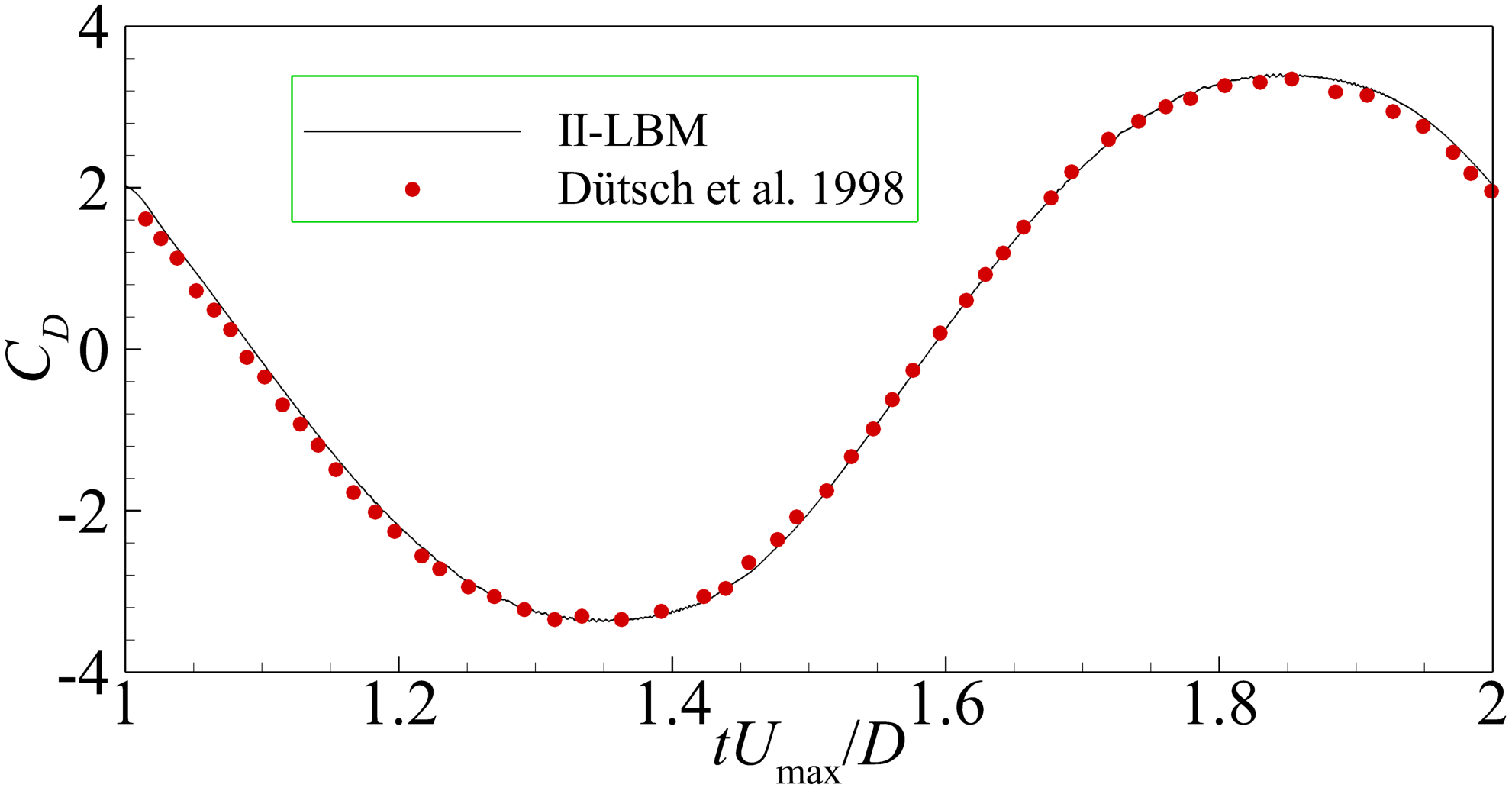}
  \caption{Comparison of drag coefficients for forced vibration of a circular cylinder in the resting fluid between the experiment of D{\"u}tsch et al.~\cite{dutsch1998low} and numerical simulation of the II-LBM with $D=80\Delta x$ at $KC=5$ and $Re=100$.}
\label{Forced_vibration}
\end{figure}

The study of forced oscillations of the structure is essential for its significance in predicting the mechanism of vortex-induced vibrations. D{\"u}tsch et al.~\cite{dutsch1998low} first studied the prescribed translation of circular cylinder by experiments and numerical simulations. In that work, a rigid circular cylinder with diameter $D$ undergoes a prescribed harmonic motion in the $x$ direction. The motion of the circular cylinder is specified with the following displacement and velocity in the $x$ direction:
\begin{equation}
\begin{aligned}
D_{\text{c}}(t)=-A\sin\left(\frac{U_{\max}t}{A}\right),\\
U_{\text{c}}(t)=-U_{\max}\cos\left(\frac{U_{\max}t}{A}\right),
\end{aligned}
\end{equation}
in which $D_{\text{c}}(t)$ and $U_{\text{c}}(t)$ are the displacement and velocity of the cylinder center in the $x$ direction, $A$ is the maximum displacement, and $U_{\max}$ is the maximum translation velocity. The Keulegan-Carpenter number and Reynolds number are $KC = \frac{2\pi A}{D}$ and $Re=\frac{U_{\max}D}{\nu}$, respectively. $KC=5$ and $Re=100$ are chosen in our present study to facilitate comparisons to D{\"u}tsch~et~al.~\cite{dutsch1998low}. The computational domain is the rectangular region $[0,55D]\times[0,35D]$, and the circular cylinder is initialized so that its center is located at the origin. Fig.~\ref{Forced_vibration} compares the drag coefficients obtained by the numerical simulation of the II-LBM with the experiment of D{\"u}tsch et al.~\cite{dutsch1998low}. The drag coefficient is defined by Eq.~(\ref{Drag_co}). Results obtained using the II-LBM are in good agreement with the prior experimental results of Eutsch~et~al.\cite{dutsch1998low}.
\floatsetup[figure]{style=plain,subcapbesideposition=top}

\subsection{Vortex-induced vibrations of a circular cylinder in the longitudinal direction}
In this section, a circular cylinder with diameter $D$ immersed in a uniform flow is considered to move freely in the $y$ direction. This case was proposed by Ahn and Kallinderis~\cite{ahn2006strongly}, and later studied by Borazjani et al.~\cite{borazjani2008curvilinear} and Bao et al.~\cite{bao2012two}. As the cylinder is placed in the fluid, fluid traction will act on the cylinder due to the no-slip boundary condition of the cylinder surface. On the other hand, the motion of the cylinder will affect the surrounding fluid. Therefore, ``two-way" interplay between the fluid and the structure is involved. The computational domain is the rectangular region $[0,40D]\times[0,20D]$, and the cylinder is initialized so that its center is located at $(10D, 10D)$. The equation to model vortex-induced vibrations of a cylinder in the longitudinal direction is
\begin{equation}
\overline{a}_y+2\zeta\left(\frac{2\pi}{U_\text{red}}\right)\overline{U}_y+\left(\frac{2\pi}{U_\text{red}}\right)^2\overline{D}_y=\frac{2C_\text{L}}{\pi \rho^*},
\label{VIV1Dof}
\end{equation}
in which $\overline{a}_y,~\overline{U}_y$, and $\overline{D}_y$ are dimensionless acceleration, velocity, and displacement of the cylinder and are normalized by $\frac{U^2}{D}, U$, and $D$, respectively. The reduced velocity is defined as $U_\text{red}=\frac{U}{f_\text{N}D}$, where $f_\text{N}=\frac{\sqrt{k_sm}}{2\pi}$ is the natural frequency of the spring. $k_\text{s}$ and $m$ are the spring constant and mass of the cylinder, respectively. $\zeta=\frac{c}{2\sqrt{k_sm}}$ is the damping ratio. The lift coefficient $C_\text{L}$ is determined in Eq.~(\ref{Lift_co}). The density ratio $\rho^*$ is the ratio between the solid density $\rho_\text{s}$ and the fluid density $\rho_\text{f}$,
\begin{equation}
\rho^*=\frac{\rho_\text{s}}{\rho_\text{f}}.
\label{density_ratio}
\end{equation}
Simulation parameters are the same as those used by Ahn and Kallinderis~\cite{ahn2006strongly}. Specifically, the density ratio and damping ratio are set to be $\frac{8}{\pi}$ and 0, respectively. The reduced velocity varies in the range of $3\leq U_\text{red}\leq 8$. The maximum oscillation amplitude of the circular cylinder for the reduced velocity from 3 to 8 are 0.075$D$, 0.551$D$, 0.520$D$, 0.459$D$, 0.370$D$, and 0.084$D$, respectively. The comparison between the present results and previous published results is shown in Fig.~\ref{Forced_vibration}.
\begin{figure}[htb!]
  \includegraphics[height=7cm]{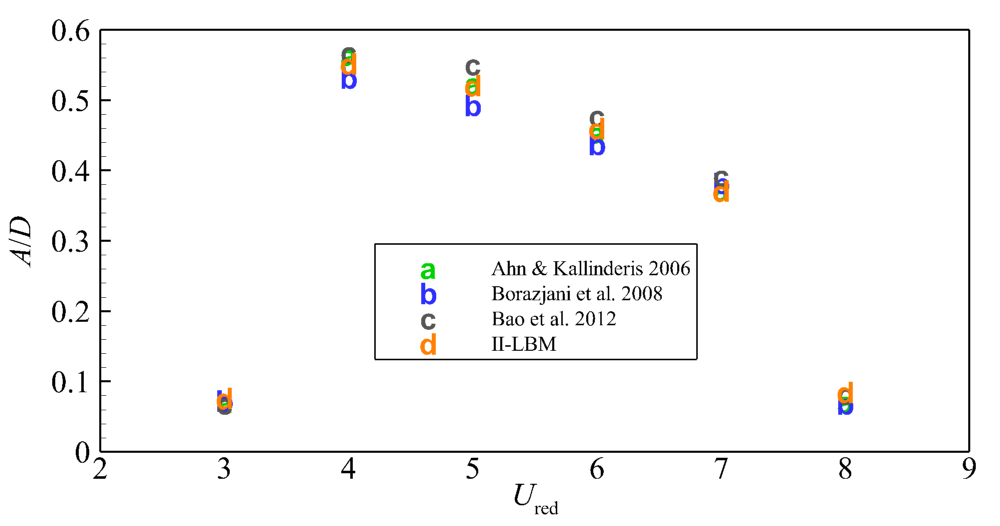}
  \caption{Comparison of the maximum displacement for vortex-induced vibrations of a circular cylinder in the $y$ direction by using the II-LBM with $D=80\Delta x$ and other published results~\cite{ahn2006strongly, borazjani2008curvilinear, bao2012two}.}
\label{VIV1DOF}
\end{figure}

\subsection{Sedimentation of an elliptical particle}
\begin{figure}[htb!]
  \includegraphics[height=6cm]{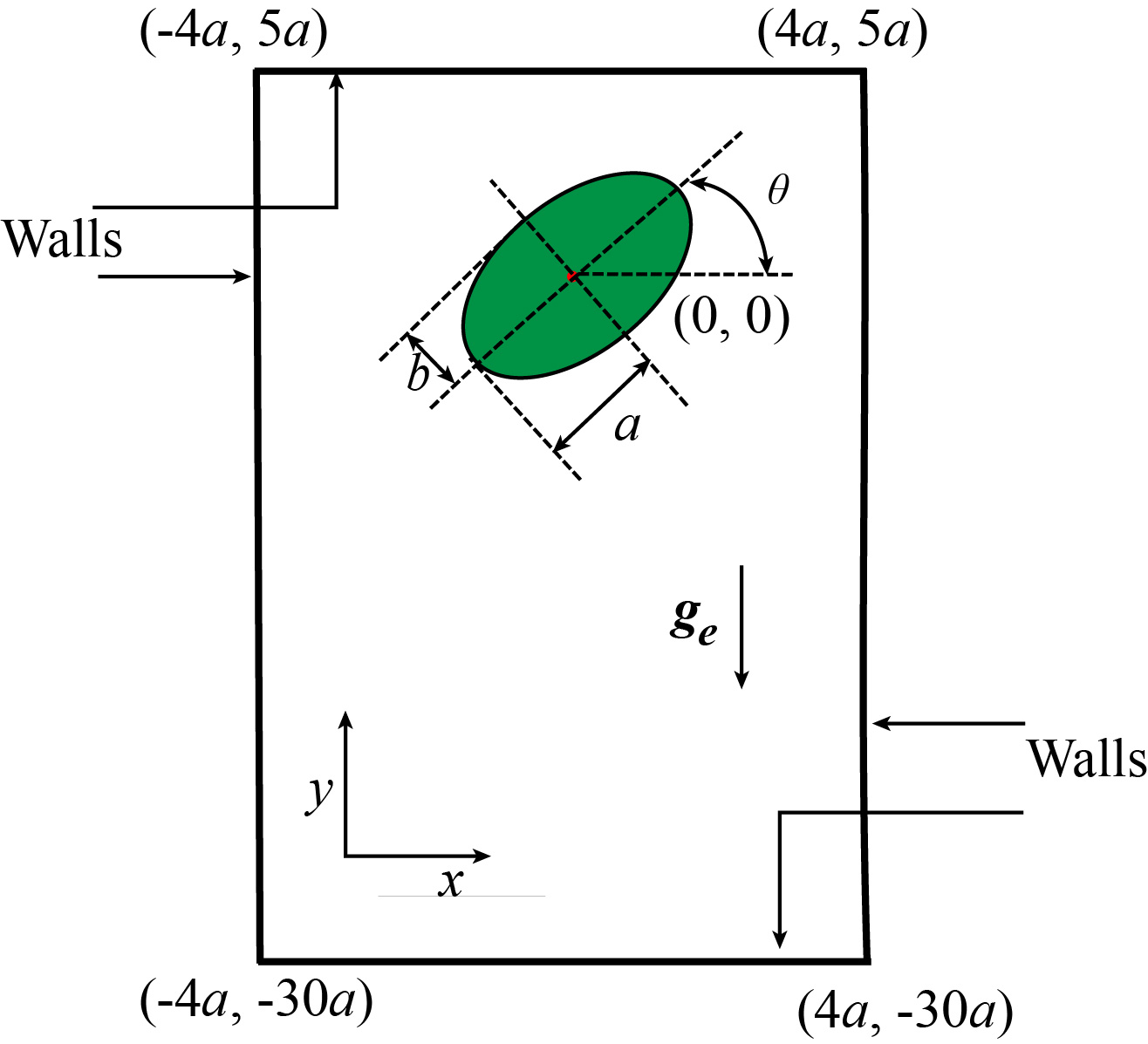}
  \caption{Schematic of sedimentation of an ellipse particle in a confined channel.}
\label{fig:Sedimentation_schematic}
\end{figure}

In this section, the sedimentation of an elliptical particle in a confined channel proposed by Xia et al.~\cite{xia2009flow} is considered. A schematic of the model is shown in Fig.~\ref{fig:Sedimentation_schematic}. When sedimenting, the ellipse particle will translate in the $x$ and $y$ directions because of gravity, pressure, and viscous forces. The particle will also rotate. Therefore, this problem involves three degrees of freedom. The computational domain is a confined box of $[-4a, 4a]\times[-30a, 5a]$, in which $a$ is the semi major axis of the ellipse particle, and the particle is initialized at (0,0) with an initial orientation angle of $\theta(t=0)=\frac{\pi}{4}$. The semi-major and semi-minor axes are taken as $a=0.025\times 10^{-2}\text{m}$ and $b=\frac{1}{2}a$, respectively. The outer boundaries are zero-velocity boundary conditions, and handled by the non-equilibrium extrapolation strategy~\cite{zhao2002non}. As in the previous section, the density ratio between the ellipse and the fluid is taken as $\rho^*$. The kinematic viscosity of the fluid is $\nu_\text{p}=0.01 \text{m}^2/\text{s}$ and gravitational acceleration is $\bm{g}_\text{e}=(0, -9.80)\,\text{m}/\text{s}^2$. 

\begin{figure}[htb!]
  \subfloat[]{\includegraphics[height=5.4cm]{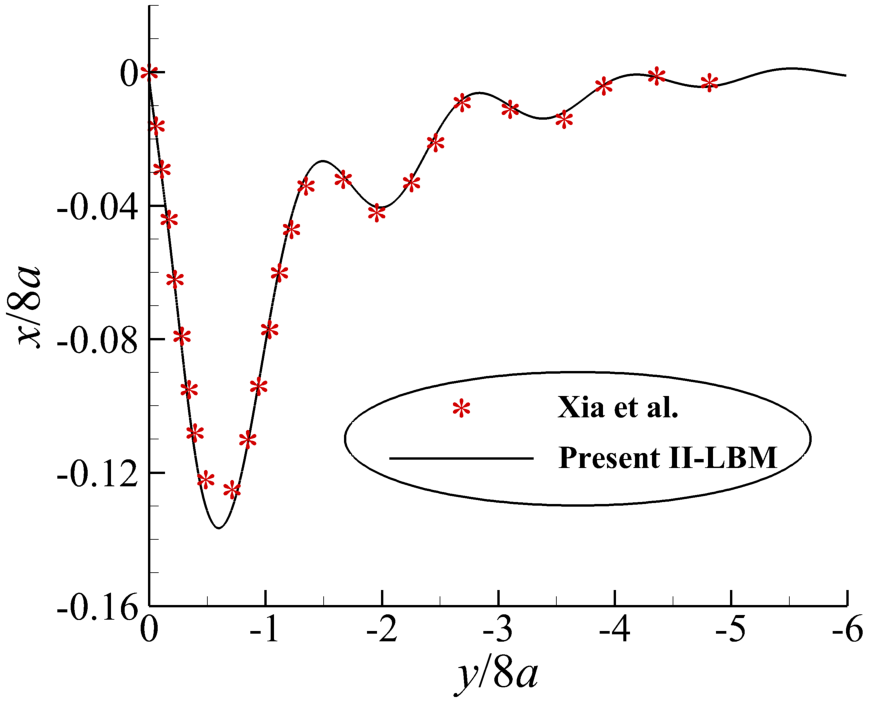}}
  \subfloat[]{\includegraphics[height=5.4cm]{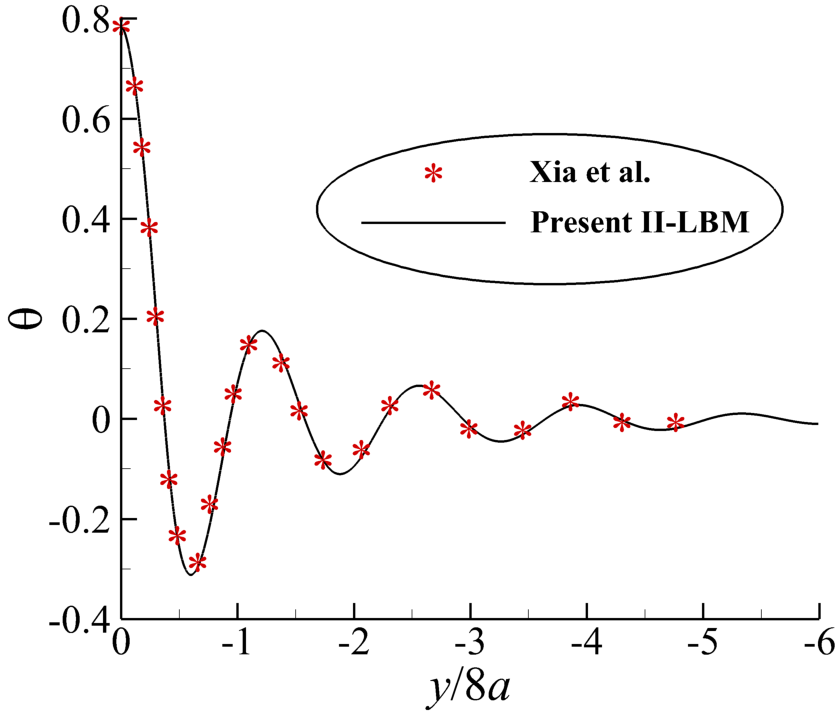}}\\
  \caption{Comparison of (a) displacement of the ellipse center in the $x$ direction and (b) angle orientation against displacement of ellipse center in the $y$ direction obtained by Xia et al. using a finite element method~\cite{xia2009flow}, and calculated by the II-LBM at $\rho^*=1.1$.}
\label{fig:Sedimentation}
\end{figure}

\begin{figure}[htb!]
  \subfloat[]{\includegraphics[height=5.4cm]{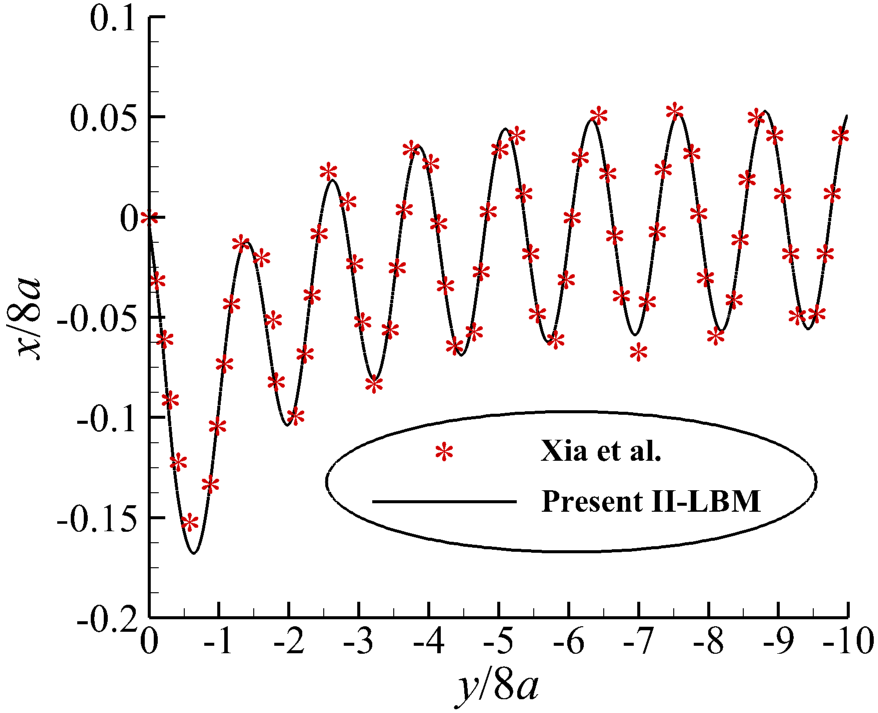}}
  \subfloat[]{\includegraphics[height=5.4cm]{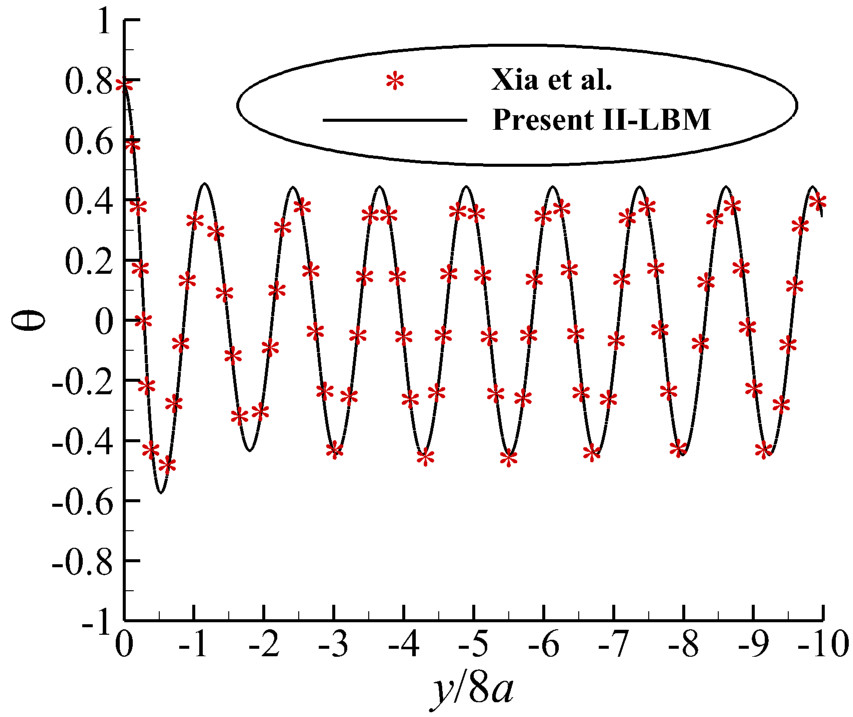}}\\
  \caption{Comparison of (a) displacement of the ellipse center in the $x$ direction and (b) angle orientation against displacement of ellipse center in the $y$ direction obtained by Xia et al. using a finite element method~\cite{xia2009flow}, and calculated by the II-LBM at $\rho^*=1.5$.}
\label{fig:Sedimentation_1.5}
\end{figure}

We first consider $\rho^*=1.1$. The grid number along the major axis is $2a=25\Delta x$, and the distance between adjacent Lagrangian points is the same as the lattice spacing. Fig.~\ref{fig:Sedimentation} shows the position of the ellipse in the $x$ direction and orientation angle of the ellipse against the position of the ellipse in the $y$ direction. Good agreement with the finite element computations of by Xia et al.~\cite{xia2009flow} is achieved. As can be seen in Fig.~\ref{fig:Sedimentation}, the amplitudes of oscillation in the displacement and orientation angle become smaller as the particle moves. When the density ratio grows to $\rho^*=1.5$, the motion of the ellipse particle in Fig.~\ref{fig:Sedimentation_1.5} calculated by using a finer grid of $2a=40\Delta x$ indicates that the ellipse approaches to a fluttering motion instead of a steady descent motion.

\section{Conclusions}
This paper determines the jump conditions for the distribution functions in the framework of the lattice Boltzmann equation and uses these conditions to develop an immersed interface-lattice Boltzmann method (II-LBM). The derived jump conditions of the distribution function can recover the pressure discontinuity induced by an immersed interface in a fluid described by the Navier-Stokes equations. Comparisons between the direct forcing IB-LBM and the II-LBM indicate that the II-LBM has higher order of accuracy for some problems and much lower errors in the velocity. Moreover,~the motion of an elastic interface shows that the II-LBM gives substantially improved volume conservation than the standard IB-LBM with both the simple external force~\cite{he1997analytic} and Guo's external force term~\cite{guo2002discrete}. The II-LBM is also applied to solve flow around a cylinder with fixed position, a circular cylinder undergoing prescribed motion in a resting fluid, vortex-induced vibrations of a circular in one degree of freedom and sedimentation of an ellipse particle. Overall, good agreement with published results is achieved for these validation cases.

As the derivation of the jump conditions of the distribution function is given in three dimensions, it would be straightforward to apply the present method to three dimensional problems.
\FloatBarrier

\section{Acknowledgements}
J.~Q.~would like to thank Nanjing University of Science and Technology for supporting the living expenses during his stay in the USA. B.~E.~G.~acknowledges research support from NSF Awards DMS 1664645, CBET 175193, OAC 1450327, OAC 1652541, and OAC 1931516. The authors thank Jae~Ho~Lee for his suggestions and comments.

\bibliography{references/New_ref}
\end{document}